\documentclass[aps,prd,onecolumn,superscriptaddress,nofootinbib]{revtex4-2}
\usepackage[utf8]{inputenc}
\usepackage{amsbsy}
\usepackage{amsmath}
\usepackage{amsfonts}
\usepackage{amssymb}
\usepackage{xcolor}
\usepackage{physics}
\usepackage{graphicx}
\usepackage{mathrsfs}
\RequirePackage[colorlinks,hyperindex]{hyperref}
\hypersetup{colorlinks=true,breaklinks=true,urlcolor=blue,linkcolor=red}
\DeclareMathOperator{\sign}{sign}

\begin{document}
\title{Missing matter  in  galaxies as a  neutrino mixing effect}

\author{Antonio Capolupo}
	\email{capolupo@sa.infn.it}
	\affiliation{Dipartimento di Fisica ``E.R. Caianiello''  Universit\`a di Salerno, and INFN -- Gruppo Collegato di Salerno, Via Giovanni Paolo II, 132, 84084 Fisciano (SA), Italy}
	\affiliation{Istituto Nazionale di Fisica Nucleare (INFN), sez. di Napoli, Gruppo Collegato di Salerno,  Italy}
		
\author{Salvatore Capozziello}
\email{capozziello@na.infn.it}
\affiliation{Dipartimento di Fisica "E. Pancini", Universit\`a degli Studi di Napoli Federico II", Via Cinthia, I-80126, Napoli, Italy}
\affiliation{Istituto Nazionale di Fisica Nucleare (INFN), sez. di Napoli, Via Cinthia 9, I-80126 Napoli, Italy}
\affiliation{Scuola Superiore Meridionale, Largo S. Marcellino 10, I-80138, Napoli, Italy}

	\author{Gabriele Pisacane}
	\email{gpisacane@unisa.it}
	\affiliation{Dipartimento di Fisica ``E.R. Caianiello''  Universit\`a di Salerno, and INFN -- Gruppo Collegato di Salerno, Via Giovanni Paolo II, 132, 84084 Fisciano (SA), Italy}
	\affiliation{Istituto Nazionale di Fisica Nucleare (INFN), sez. di Napoli, Gruppo Collegato di Salerno, Italy}
	
		\author{Aniello Quaranta}
	\email{anquaranta@unisa.it}
	\affiliation{Dipartimento di Fisica ``E.R. Caianiello''  Universit\`a di Salerno, and INFN -- Gruppo Collegato di Salerno, Via Giovanni Paolo II, 132, 84084 Fisciano (SA), Italy}
	\affiliation{Istituto Nazionale di Fisica Nucleare (INFN), sez. di Napoli, Gruppo Collegato di Salerno, Italy}.

\begin{abstract}

We show that, in the framework of quantum field theory in curved spacetime, the semiclassical energy-momentum tensor of  the neutrino flavor vacuum fulfills the equation of state  of dust and cold dark matter.
We consider spherically symmetric spacetimes, and we demonstrate that, within the weak field approximation, the flavor vacuum contributes as a Yukawa correction to the Newtonian potential.  This corrected potential may account for the flat rotation curves of spiral galaxies. In this perspective, neutrino mixing could contribute to dark  matter.

\end{abstract}

\maketitle

\section{Introduction}
 In the framework of $\mathrm{\Lambda CDM}$ model,  the today cosmology can account for the observed universe at the price of introducing unknown ingredients, such as the cosmological constant $\Lambda$ or a similarly behaving dark energy to explain the accelerated expansion \cite{DE1,DE2,DE3,DE4,DE5,DE6,DE7}, and  non-baryonic dark matter \cite{Dm1,Dm2,Dm3,Dm4,Dm5,Dm6,Dm7,Dm8,Dm9,Dm10,Dm11,Dm12,Dm13,Dm14,Dm15,Dm16,Dm17,Dm18} to justify the flat rotation curves of spiral galaxies and the gravitational stability of large scale structures. Remarkably, the dark components, whose composition and origin represent fundamental open issues, constitute the overwhelming majority of the total energy density of the universe. Several beyond the standard model (BSM) theories suggest an explanation of dark matter in terms of new hypothetical particles: axions and axion-like particles \cite{Axion1,Axion2,Axion3,Axion4,Axion5,Axion6,Axion7,Axion8,Axion9,Axion10,Axion11,Gan}, mirror matter \cite{MM1,MM2,MM3,MM4,MM5,MM6,MM7}, new gauge bosons  \cite{R1,R2,R3,R4,R5},  supersymmetric particles \cite{SSDM} and sterile neutrinos \cite{SN}, to name a few. No convincing experimental evidence has been found for these particles as yet. Another possible interpretation of dark matter comes from extended theories of gravity \cite{ETG1,ETG2,ETG3} or some alternative to General Relativity \cite{Cai}.

 On the other hand it is established that massive neutrinos \cite{N1,N2,N3,N4,N5,N6,N7,N8,N9,N10,N11,N12,N13,N14} play an important role in astrophysical and cosmological contexts \cite{CN1,CN2,CN3,CN4,CN5,CN6,CN7,CN8,CN9,CN10,FV1,FV2,FV3}. Representing the most direct challenge to the standard model of particles, neutrino mixing has numerous phenomenological implications. The quantum field theory (QFT) of neutrino mixing \cite{FV1,FV2,FV3} involves a non-trivial vacuum state, the flavor vacuum. In virtue of its condensate structure, the latter may represent an additional, unconventional, gravitating source, emerging directly from QFT and comparable to other known condensation phenomena \cite{CND1,CND2,CND3,CND4,CND5,CND5,CND6,Capolupo2016,CND7,CND8,CND9,CND10,CND11,CND12}. Although it has been hinted that the flavor vacuum may be related to the dark components of the universe \cite{FV1,FV2,FV3,CND2,CND3,Capolupo2016}, in particular for cosmological metrics \cite{FV2,FV3}, a general result in this respect was still lacking.

 In the present work, employing QFT in curved space, we establish a general result on the energy-momentum tensor associated to the flavor vacuum of neutrino mixing, valid for any sufficiently regular spacetime. We show that under very weak requirements, the energy-momentum tensor associated to the flavor vacuum acquires a perfect fluid form, with equation of state typical of dust and cold dark matter. We therefore generalize and recover the results previously obtained in flat space \cite{Capolupo2016} and in a Friedmann-Lema\^itre-Robertson-Walker (FLRW)  metric with de Sitter evolution \cite{FV2,FV3}.

 We then turn to the paramount case of spherically symmetric spacetimes, developing the quantization of flavor fields and deriving the cold dark matter equation of state for the energy-momentum tensor associated to the flavor vacuum. We demonstrate that, within the weak field approximation, the flavor vacuum contributes as a Yukawa correction to the Newtonian potential. It is then discussed how this may account for the flat rotation curves of spiral galaxies, reinforcing the interpretation of the flavor vacuum as a possible dark matter component.
 In other words, neutrino mixing could be considered as a sort of MOND effect generating corrections to the gravitational potential acting at galactic scales (see e.g. \cite{Milgrom, Milgrom2, Benisty}).

 The paper is structured as follows: Section II presents a brief recap of the QFT of flavor fields in curved space. In Section III we establish, on purely algebraic grounds, the general properties of the Vacuum Expectation Value (VEV) of the energy momentum tensor on the flavor vacuum, showing that it attains the perfect fluid form in many cases of interest. Section IV is devoted to the detailed analysis of the flavor vacuum in static spherically symmetric metrics and culminates in the derivation of the induced gravitational potential within the weak field approximation. In Section V, the result is adopted to fit   a sample of spiral galaxies whose main dynamical characteristic is the presence of dark matter haloes. Finally, we derive the baryonic Tully-Fisher relation. Section VI is devoted to discussion and  conclusions.

\section{A summary of Flavor Field Quantization}

Let us  start by briefly recalling the basic aspects of flavor field quantization in curved spacetime. A more comprehensive treatment can be found in \cite{FV1,FV2,FV3}. We limit ourselves to two flavors for the sake of simplicity. The introduction of the third flavor complicates the analysis and introduces some novel features, such as the Dirac CP violation, nonetheless without altering the fundamental properties of the theory discussed here, nor the ensuing results shown in the upcoming sections. The action for the two free Dirac fields $\nu_1 (x), \nu_2 (x)$, the neutrino fields with  masses $M_1, M_2$, is
\begin{equation}\label{Eq.:Action}
 S = \sum_{L=1,2} \int d^4 x \ \sqrt{-g} \left[ \frac{i}{2} \left(\bar{\nu}_L \tilde{\gamma}^{\mu} D_{\mu} \nu_L - D_{\mu} \bar{\nu}_L \tilde{\gamma}^{\mu} \nu_L \right)-M_L \bar{\nu}_L \nu_L \right] \ .
\end{equation}
Here $\sqrt{-g}$ is the determinant of the metric tensor $g_{\mu \nu}$, $\bar{\nu} = \nu^{\dagger} \gamma^0$ and a choice of tetrads $e^{A}_{\mu}$ is understood, so that the curved space gamma matrices $\tilde{\gamma}^{\mu}$ and the spinorial covariant derivatives $D_{\mu}$ are respectively given as $\tilde{\gamma}^{\mu} = e^{\mu}_A \gamma^A$ and $D_{\mu} = \partial_{\mu} + \Gamma_{\mu}$ ($D_{\mu}= \partial_{\mu} - \Gamma_{\mu}$ when acting on adjoint spinors $\bar{\nu}$), with $\Gamma_{\mu} = \frac{1}{8}\omega_\mu^{AB} \left[ \gamma_A, \gamma_B \right]$. Note that uppercase latin indices are here used for the Minkowskian indices and the spin connection is given as usual as $\omega_{\mu}^{AB} = e^A_{\rho} \Gamma^{\rho}_{\nu \mu} e^{\nu B} + e^{A}_{\rho} \partial_{\mu} e^{\rho B}$. We assume that the underlying manifold is globally hyperbolic, foliated by the Cauchy surfaces $\Sigma_{\tau}$. It is  convenient to choose the foliation parameter $\tau$ corresponding to the time coordinate $x^0$. The inner product for any two spinors $U,V$ at time $\tau$ is then
\begin{equation}\label{Eq.:InnerProduct}
 \left( U, V \right)_{\tau} = \int_{\Sigma_{\tau}} d \Sigma_{\mu} \sqrt{-g} \bar{U} \tilde{\gamma}^\mu V \ .
\end{equation}
Given any complete\footnote{With respect to the inner product of Eq. \eqref{Eq.:InnerProduct}, see \cite{FV1,FV2,FV3} for details.} set of solutions $\left \lbrace U_{L,q,s}, V_{L,q,s}\right \rbrace$ to the Dirac equations
\begin{equation}\label{Eq.:DiracEquations}
 \left( i \tilde{\gamma}^{\mu}D_{\mu} - M_L \right) \nu_L = 0 \ ,
\end{equation}
with $q$ and $s$ generalized momentum and spin indices, the mass fields are expanded as usual
\begin{equation}\label{Eq.:FreeFieldExpansion}
 \nu_L (x) = \sum_s \int d^3 q \left(a_{L,q,s} U_{L,q,s} (x) + b^{\dagger}_{L,q,s} V_{L,q,s}(x) \right) \ .
\end{equation}
In virtue of the expansion \eqref{Eq.:FreeFieldExpansion},  one defines the mass vacuum $\ket{0}$ according to $a_{L,q,s}\ket{0} = 0 = b_{L,q,s} \ket{0}$. The physical (flavor) fields are obtained via a rotation from the mass fields: $\nu_e (x) = \cos \Theta \nu_1 (x) + \sin \Theta \nu_2 (x)$ and $\nu_{\mu} (x) = \cos \Theta \nu_2 (x) - \sin \Theta \nu_1 (x)$, where $\Theta$ denotes the two flavor mixing angle. In the language of Quantum Field Theory, the rotation of the fields is achieved by means of the mixing generator $\mathcal{J}_{\Theta} (\tau) = \exp \left\lbrace \Theta \left[\left(\nu_1,\nu_2 \right)_{\tau} - \left(\nu_2,\nu_1 \right)_{\tau}\right] \right\rbrace$
\begin{equation}\label{Eq.:MixingGenerator}
  \begin{pmatrix}
\nu_e (x) \\ \nu_\mu(x) \end{pmatrix} =\mathcal{J}_{\Theta}^{-1} (\tau) \begin{pmatrix}
\nu_1 (x) \\ \nu_2(x)
\end{pmatrix}\mathcal{J}_{\Theta} (\tau) = \begin{pmatrix} \cos \Theta & \sin \Theta \\ - \sin \Theta & \cos \Theta \end{pmatrix}
\begin{pmatrix}
\nu_1 (x) \\ \nu_2(x)
\end{pmatrix} \ .
\end{equation}
Here it is understood that all the fields are evaluated at $x^0 = \tau$. Notice, importantly, that the mixing generator depends explicitly on time but has no dependence on the spacelike coordinates. Note also that, by definition, it is  $\mathcal{J}^{-1}_{\Theta}(\tau) = \mathcal{J}_{-\Theta}(\tau)$. The action of the generator on the annihilators leads to a Bogoliubov transformation nested into a rotation, thus defining the flavor operators
\begin{eqnarray}\label{Eq.:FermionOperators}
 \nonumber a_{e,q,s}(\tau) &=& \mathcal{J}_{\Theta}^{-1} (\tau) a_{1,q,s} \mathcal{J}_{\Theta} (\tau) = \cos \Theta \ a_{1,q,s} + \sin \Theta \sum_{p,r} \left(\Xi^*_{q,s;p,r} (\tau) a_{2,p,r} + \Omega_{q,s;p,r} (\tau)b^{\dagger}_{2,p,r} \right) \\
 \nonumber a_{\mu,q,s}(\tau) &=& \mathcal{J}_{\Theta}^{-1} (\tau) a_{2,q,s} \mathcal{J}_{\Theta} (\tau) = \cos \Theta \ a_{2,q,s} - \sin \Theta \sum_{p,r} \left(\Xi_{q,s;p,r} (\tau) a_{1,p,r} - \Omega_{q,s;p,r} (\tau)b^{\dagger}_{1,p,r} \right) \\
  \nonumber b_{e,q,s}(\tau) &=& \mathcal{J}_{\Theta}^{-1} (\tau) b_{1,q,s} \mathcal{J}_{\Theta} (\tau) = \cos \Theta \ b_{1,q,s} + \sin \Theta \sum_{p,r} \left(\Xi^*_{q,s;p,r} (\tau) b_{2,p,r} - \Omega_{q,s;p,r} (\tau)a^{\dagger}_{2,p,r} \right) \\
  b_{\mu,q,s}(\tau) &=& \mathcal{J}_{\Theta}^{-1} (\tau) b_{2,q,s} \mathcal{J}_{\Theta} (\tau) = \cos \Theta \ b_{2,q,s} - \sin \Theta \sum_{p,r} \left(\Xi_{q,s;p,r} (\tau) b_{1,p,r} + \Omega_{q,s;p,r} (\tau)a^{\dagger}_{2,p,r} \right) \ .
\end{eqnarray}
In the above expressions,  the generalized sum $\sum_{p,r}$ involves an integral on the momentum index $p$ if this is continuous and we have introduced the Bogoliubov coefficients of mixing $\Xi_{q,s;p,r} (\tau) = \left(U_{2,q,s}, U_{1,p,r} \right)_{\tau}$ and $\Omega_{q,s;p,r} (\tau) = \left(U_{1, p,r}, V_{2,q,s} \right)_{\tau}$. The operators \eqref{Eq.:FermionOperators} annihilate the time dependent flavor vacuum $\ket{0_F (\tau)} = \mathcal{J}^{-1}_{\Theta} \ket{0}$. This new vacuum belongs to a unitarily inequivalent representation with respect to the mass vacuum and has the structure of a condensate of particle-antiparticle pairs with definite masses. Because of the condensate structure, the energy-momentum content of the flavor vacuum is non-trivial. The energy-momentum tensor, derived from the action \eqref{Eq.:Action}, is
\begin{equation}\label{Eq.:EnergyMomentumTensor}
 T_{\mu \rho} (x) = \frac{i}{2}\sum_{L=1,2} \left[ \bar{\nu}_L \tilde{\gamma}_{(\mu} D_{\rho)} \nu_L - D_{(\mu} \bar{\nu}_L \tilde{\gamma}_{\rho)} \nu_L\right] \ .
\end{equation}
The instantaneous energy-momentum content of the flavor vacuum is
catched by the expectation value
\begin{equation}\label{Eq.:VEV}
 \mathbb{T}_{\mu \rho} ( x) = \bra{0_F (\tau)} : T_{\mu \rho} (x) : \ket{0_F (\tau)}
\end{equation}
where normal ordering $: .. :$ is with respect to the mass vacuum\footnote{This isolates the pure mixing contribution, see \cite{FV2,FV3}.} $\ket{0}$. The vacuum expectation value $\mathbb{T}_{\mu \rho}$ is the central object of our analysis.
Closely related to Eq. \eqref{Eq.:VEV}, it  is the non-synchronous expectation value
\begin{equation}\label{Eq.:VEVNS}
 \mathbb{T}_{\mu \rho} (\tau_0, x) = \bra{0_F (\tau_0)} : T_{\mu \rho} (x) : \ket{0_F (\tau_0)} \ ,
\end{equation}
that reduces to Eq. \eqref{Eq.:VEV} when $\tau_0 = x^0$. With these considerations in mind, let us now calculate the VEV of the energy-momentum tensor.

\section{THE VACUUM EXPECTATION VALUE  OF THE ENERGY-MOMENTUM TENSOR}

We now establish the general properties of $\mathbb{T}_{\mu \rho} (x)$ and $\mathbb{T}_{\mu \rho} (\tau_0,x)$. Let $\nabla$ be any linear operator, possibly involving spatial derivatives (but no time derivatives), with any tensor structure. For instance $\nabla \equiv \tilde{\gamma}_i D_j$ with $i,j = 1,2,3$ denoting spatial indices. Any bilinear combination of the form
\begin{equation}\label{Eq.:bilinear}
 \bar{\nu}_1 (x) (\nabla) \nu_1(x) + \bar{\nu}_2(x) (\nabla) \nu_2(x)
\end{equation}
is invariant under the action of the mixing generator. One indeed has
\begin{eqnarray}\label{Eq.:InvarianceProof}
\nonumber && \mathcal{J}_{\Theta}^{-1} (\tau) \bar{\nu}_1 (x) (\nabla) \nu_1(x) \mathcal{J}_{\Theta}(\tau) + \mathcal{J}_{\Theta}(\tau)\bar{\nu}_2(x) (\nabla) \nu_2(x)\mathcal{J}^{-1}_{\Theta}(\tau) =
\\[0.2cm]
\nonumber && \mathcal{J}_{\Theta}^{-1} (\tau) \bar{\nu}_1 (x) \mathcal{J}_{\Theta} (\tau) \mathcal{J}_{\Theta}^{-1} (\tau)(\nabla) \nu_1(x) \mathcal{J}_{\Theta}(\tau) + \mathcal{J}_{\Theta}(\tau)\bar{\nu}_2(x) \mathcal{J}_{\Theta} (\tau) \mathcal{J}_{\Theta}^{-1} (\tau) (\nabla) \nu_2(x)\mathcal{J}^{-1}_{\Theta}(\tau) =
\\[0.2cm]
\nonumber && \mathcal{J}_{\Theta}^{-1} (\tau) \bar{\nu}_1 (x) \mathcal{J}_{\Theta} (\tau)  (\nabla) \mathcal{J}_{\Theta}^{-1} (\tau) \nu_1(x) \mathcal{J}_{\Theta}(\tau) + \mathcal{J}_{\Theta}(\tau)\bar{\nu}_2(x) \mathcal{J}_{\Theta} (\tau)   (\nabla) \mathcal{J}_{\Theta}^{-1}(\tau) \nu_2(x)\mathcal{J}^{-1}_{\Theta}(\tau) =
\\[0.2cm]
\nonumber && \left(\cos \Theta \bar{\nu}_1(x) + \sin \Theta \bar{\nu}_2(x) \right) (\nabla) \left(\cos \Theta \nu_1(x) + \sin \Theta \nu_2(x) \right) + \left(\cos \Theta \bar{\nu}_2(x) - \sin \Theta \bar{\nu}_1(x) \right) (\nabla) \left(\cos \Theta \nu_2(x) - \sin \Theta \nu_1(x) \right) =
\\[0.2cm]
\nonumber && \left(\cos^2 \Theta + \sin^2 \Theta \right) \left( \bar{\nu}_1 (x) (\nabla) \nu_1(x) + \bar{\nu}_2(x)(\nabla) \nu_2 (x) \right)
\\
\nonumber && + \cos \Theta \sin \Theta \left(\bar{\nu}_1 (x) (\nabla) \nu_2(x) + \bar{\nu}_2 (x) (\nabla) \nu_1(x) - \bar{\nu}_2 (x) (\nabla) \nu_1(x) - \bar{\nu}_1 (x) (\nabla) \nu_2(x)\right) =
\\[0.2cm]
&& \bar{\nu}_1 (x) (\nabla) \nu_1(x) + \bar{\nu}_2(x) (\nabla) \nu_2(x) \ .
\end{eqnarray}
In the second line,  we have inserted the identity $\mathcal{J}_{\Theta}(\tau)\mathcal{J}^{-1}_{\Theta}(\tau) $. The third line descends from $\left[ \nabla, \mathcal{J}_{\theta} (\tau) \right] = 0$, which holds because $\nabla$ does not involve time derivatives and the mixing generator depends only on time\footnote{It is further clear that the mixing generator is a scalar with respect to the spacetime algebra and the Clifford algebra of $\gamma$ matrices.}. This is the crucial step: bilinears that contain time derivatives are not left invariant by the action of the mixing generator. In the fourth line,  we have simply substituted Eq. \eqref{Eq.:MixingGenerator}. All the equalities are understood to hold on the surface $\Sigma_\tau$, i.e. for $x^0=\tau$. We remark that the same argument is valid for bilinears of the form $
 \bar{\nu}_1 (x) (\overleftarrow{\nabla})  \nu_1(x) + \bar{\nu}_2(x) (\overleftarrow{\nabla}) \nu_2(x)
$, in which the eventual derivatives act on the left, e.g. $\sum_{L=1,2} D_j \bar{\nu}_L \tilde{\gamma}_i \nu_L$. In addition, since Eq. \eqref{Eq.:InvarianceProof} is true for any $\Theta$, it holds also for $- \Theta$, and therefore even if the mixing generator is replaced by its inverse. It is an immediate consequence of Eq. \eqref{Eq.:InvarianceProof} that all the spatial components of $\mathbb{T}_{\mu \nu}(x)$ and $\mathbb{T}_{\mu \nu}(\tau_0,x)$, vanish, that is
\begin{equation}\label{Eq.:NoSpatialComponents}
 \mathbb{T}_{jk} (x) = 0 = \mathbb{T}_{jk} (\tau_0,x) \ \ \ \ \ \ \ \forall j, k = 1,2,3 \ .
\end{equation}
To prove this, let us start with
\begin{equation}\label{Eq.:FirstStep}
 \mathbb{T}_{jk} (x) = \bra{0_F(\tau)} : T_{jk} : \ket{0_F (\tau)} = \bra{0_F(\tau)}  T_{jk}  \ket{0_F (\tau)} - C_{jk} \ .
\end{equation}
Here we have introduced the normal ordering constant $C_{jk}= \bra{0} T_{jk} \ket{0}$. Using the relation between the flavor vacuum and the mass vacuum,  we arrive at
\begin{equation}\label{Eq.:SecondStep}
 \bra{0_F(\tau)} T_{jk} \ket{0_F(\tau)} = \bra{0} \mathcal{J}_{\Theta}(\tau) T_{jk} \mathcal{J}_{\Theta}^{-1} (\tau) \ket{0} =  \bra{0} \mathcal{J}^{-1}_{-\Theta}(\tau) T_{jk} \mathcal{J}_{-\Theta} (\tau) \ket{0} = \bra{0} T_{jk} \ket{0} \ ,
\end{equation}
where the last equality follows from Eq. \eqref{Eq.:InvarianceProof} and noting that $T_{jk}$, as shown in Eq. \eqref{Eq.:EnergyMomentumTensor}, is the sum of bilinears of the form \eqref{Eq.:bilinear} (and their left-acting relatives) that contain only spatial derivatives. Substituting Eq. \eqref{Eq.:SecondStep} in \eqref{Eq.:FirstStep},  the assertion \eqref{Eq.:NoSpatialComponents} is proved.

The same argument, of course, cannot be applied to the components that involve time derivatives $\mathbb{T}_{00}$ and $\mathbb{T}_{0i}$, which are, in general,  different  from zero. The remaining off-diagonal components $\mathbb{T}_{0i}$ are proven to vanish in most cases of interest (see for instance \cite{FV2} for FLRW metrics, and the section below for spherically symmetric spacetimes), so that the only surviving component is $\mathbb{T}_{00}$. In these cases,  the energy-momentum tensor of the flavor vacuum attains a perfect fluid form with vanishing pressure and it is therefore characterized by the equation  of state of dust and cold dark matter:
\begin{equation}\label{Eq.:EquationOfState}
p = 0 =  w \rho \leftrightarrow \mathbb{T}_{jj} = 0 = w \mathbb{T}_{00} \rightarrow w = 0  \ ,
\end{equation}
where we have introduced the energy density $\rho = \mathbb{T}_{00}$ and the pressure $\mathbb{T}_{jj} = p = 0$. Here $j$ can be any of $1,2,3$.

\section{Spherical Symmetry}

It was proven in \cite{FV2} that, for a FLRW metric, the energy-momentum tensor associated to the flavor vacuum is diagonal $\mathbb{T}_{0i} = 0$. This fact, in conjunction with the general result of Eq. \eqref{Eq.:NoSpatialComponents} allows us to immediately deduce that $\mathbb{T}_{\mu \nu}$ corresponds to the energy-momentum tensor of a pressureless perfect fluid in FLRW metrics, at any time and regardless of the specific scale factor $C(t)$. This constitutes a significant generalization of the results derived in \cite{FV2}. From a cosmological  point of view,  the flavor vacuum may then play the role of a cold dark matter component.

Another class of metrics of interest, relevant in astrophysics and for the description of galaxies, is that of static spherically symmetric spacetimes. The celebrated Schwarzschild solution belongs to this class. In isotropic coordinates, we can write the metric as (see for instance \cite{Egorov}):
\begin{equation}\label{Eq.:SphericalMetric}
 ds^2 = f(R) dt^2 - g(R) \left(dx^2 + dy^2 + dz^2 \right)
\end{equation}
with $R = \sqrt{x^2 + y^2 + z^2}$ and $f(R),g(R)$ arbitrary, sufficiently regular, functions of $R$. The following analysis can be extended, at the cost of some minor algebraic complications, to the case of time dependent spherically symmetric metrics $f(R,t), g(R,t)$. We stick to the static case for simplicity. The natural tetrad is provided by $e^{A}_0 = \sqrt{f(R)}\delta^A_0$ and $e^{A}_j = \sqrt{g(R)} \delta^A_j$ for $j=1,2,3$. The non-vanishing Christoffel symbols are
\begin{equation}\label{Eq.:Christoffel}
\Gamma^0_{0j} = \frac{f'(R)}{2f(R)}\frac{x^j}{R} \ ; \ \ \Gamma^{j}_{00} = \frac{f'(R)}{2g(R)} \frac{x^j}{R} \ ; \\ \ \Gamma^{j}_{kj} = \frac{g'(R)}{2g(R)} \frac{x^k}{R} \ k \neq j \ ; \ \ \ \ \ \ \ \ \Gamma^{j}_{kk} = -\frac{g'(R)}{2g(R)} \frac{x^j}{R} \ k \neq j \ ; \ \ \ \Gamma^{j}_{jj} = \frac{g'(R)}{2g(R)} \frac{x^j}{R} \ ,
\end{equation}
with  prime denoting the derivative with respect to $R$. It is straightforward to compute the spin connections, which lead to
\begin{equation}
 \Gamma_{0} = \frac{1}{8} \frac{f'(R)}{\sqrt{f(R)g(R)}} \sum_{j=1,2,3} \frac{x^j}{R}\left[\gamma^0, \gamma^j \right] \ ; \ \ \ \Gamma_j = \frac{g'(R)}{8g(R)} \sum_{k \neq j} \frac{x^k}{R} \left[\gamma^k, \gamma^j \right] \ .
\end{equation}
The contraction $\tilde{\gamma}^{\mu} \Gamma_{\mu}$, needed for the Dirac equation, reads neatly
\begin{equation}
 \tilde{\gamma}^{\mu}\Gamma_{\mu} = T(R) \ \hat{\pmb{n}} \cdot \pmb{\gamma} \ \  \ , \ \ \ \ \ \ \ \ \ \ \ \ \ T(R) := \frac{f'(R)}{4f(R)\sqrt{g(R)}} + \frac{g'(R)}{2g(R) \sqrt{g(R)}} \ .
\end{equation}
In the above equation, we have introduced the spatial vectors $\pmb{\gamma}$ and $\hat{\pmb{n}} = \frac{\pmb{x}}{R}$, along with the Euclidean $3$-d product $\cdot$. Finally the Dirac equation for the mass $M_L$ reads
\begin{equation}\label{Eq.:SphericalDiracEquation}
 \frac{i}{\sqrt{f(R)}} \gamma^0 \partial_0 u_L + \frac{i}{\sqrt{g(R)}} \pmb{\gamma} \cdot \pmb{\nabla} u_L + i T(R) \ \hat{\pmb{n}} \cdot \pmb{\gamma} u_L - M_L u_L = 0 \ ,
\end{equation}
where $\pmb{\nabla}$ is the usual spatial gradient. The spherical symmetry suggests the following ansatz for the positive energy solutions:
\begin{equation}\label{Eq.:SolutionAnsatz}
 u_{L, E_L,j,\kappa, m_j} (t, \pmb{x}) = e^{-iE_L t} \begin{pmatrix}
   \Phi_{L,E_L,j,\kappa}(R) \chi_{\kappa, m_j} (\hat{\pmb{n}}) \\ i \Psi_{L, E_L,j,\kappa}(R) \chi_{-\kappa, m_j}(\hat{\pmb{n}})
\end{pmatrix} \ .
\end{equation}
 Here, $j$ denotes the eigenvalues of the total angular momentum operator $J^2$ and $m_j$ the eigenvalues of its $z$ component $J_z$. For each $j= \frac{1}{2}, \frac{3}{2}, ...$ there are two values for the spin-orbit quantum number $\kappa = \pm (j + \frac{1}{2})$, so that
$\kappa \in \mathbb{Z} \smallsetminus \left\lbrace 0 \right\rbrace$. The two-component spherical spinors $\chi_{\kappa ,m_j}$ are eigenspinors of $J^2, J_z$ and $K$ respectively with eigenvalues $j,m_j,\kappa$. These spinors satisfy a number of useful properties \cite{SphericalSpinors}, some of which we quote here for  the reader convenience:
\begin{eqnarray}\label{Eq.:SphericalSpinorProperties}
 \nonumber \left(\hat{\pmb{n}} \cdot \pmb{\sigma}\right) \chi_{\kappa,m_j} &=& - \chi_{-\kappa,m_j} \ ; \ \ \ \ \ \left(\pmb{\sigma} \cdot \pmb{\nabla} \right) \left( H(R) \chi_{\kappa, m_j} \right) = - \left(\partial_{R} + \frac{\kappa + 1}{R} \right)H(R) \chi_{-\kappa, m_j} \\
 \sigma_z \chi_{\kappa m_j} &=& G_{\kappa, m_j} \chi_{\kappa, m_j} + H_{\kappa, m_j} \chi_{-\kappa -1, m_j} := - \frac{2m_j}{2\kappa + 1}\chi_{\kappa, m_j} - 2 \frac{\sqrt{\left(\kappa + \frac{1}{2}\right)^2 - m_j^2}}{|2\kappa + 1 |}\chi_{-\kappa -1, m_j} \ .
\end{eqnarray}
We have introduced the shorthand notation $G_{\kappa, m_j}, H_{\kappa,m_j}$ for later convenience and $H(R)$ denotes an arbitrary function of $R$, whereas $\pmb{\sigma}$ indicates the vector of Pauli matrices. By employing the Dirac representation for the flat gamma matrices and using the properties of Eq. \eqref{Eq.:SphericalSpinorProperties},  the Dirac equation separates fully, yielding a system of equations for the radial functions
\begin{eqnarray}\label{Eq.:RadialSystem}
 \nonumber \left(\partial_R + \frac{1-\kappa}{R} \right)\Psi_{L,E_L,j,\kappa}(R) &=& \left(M_L \sqrt{g(R)} - E_L\sqrt{\frac{g(R)}{f(R)}} \right) \Phi_{L,E_L,j,\kappa}(R) - \sqrt{g(R)}T(R) \Psi_{L,E_L,j,\kappa}(R)\,, \\
 \left(\partial_R + \frac{1+\kappa}{R} \right)\Phi_{L,E_L,j,\kappa}(R) &=& \left(M_L \sqrt{g(R)} + E_L\sqrt{\frac{g(R)}{f(R)}} \right) \Psi_{L,E_L,j,\kappa}(R) - \sqrt{g(R)}T(R) \Phi_{L,E_L,j,\kappa}(R) \ .
\end{eqnarray}
The system can be solved explicitly once the metric functions $g(R), f(R)$ are specified. The radial functions determine also the corresponding antiparticle solutions, that are obtained by means of a simple charge conjugation from Eq. \eqref{Eq.:SolutionAnsatz}:
\begin{equation}\label{Eq.:AntiparticleSolution}
 v_{L,E_L,j,\kappa,m_j} (t, \pmb{x}) =  \mathrm{sign}(\kappa) (-1)^{m_j + \frac{3}{2}} e^{iE_L t}\begin{pmatrix}
  i \Psi_{L,E_L,j,\kappa}(R) \chi_{-\kappa, - m_j} (\hat{\pmb{n}}) \\
  -\Phi_{L,E_L,j,\kappa} (R) \chi_{\kappa, -m_j} (\hat{\pmb{n}})
                                                                                    \end{pmatrix} \ .
\end{equation}
In principle, the radial functions in the above equation should carry a complex conjugation. Yet a glance at the system \eqref{Eq.:RadialSystem} reveals that all the coefficients are real, so that the radial functions can always be chosen to be real. Inspired by the Minkowskian limit, we shift the labelling of the modes from $E_L$ to $p_L = \sqrt{E_L^2 - M_L^2}$, that plays the role of a radial momentum for the species $L$. Note that this shift is well defined for $E_L \geq M_L$. When $E_L < M_L$ we set $q_L= \sqrt{M_L^2 - E_L^2}$, but its interpretation in terms of a radial momentum is no longer sensible\footnote{There is no such modes in the Minkowskian limit, where $E_L \geq M_L$ always.}. The inner product of Eq. \eqref{Eq.:InnerProduct}, specified for the metric at hand,  is
\begin{equation}
 \left(U,V\right)_t = \int_0^{\infty} dR g^{\frac{3}{2}}(R) R^2 \int_{4\pi} d \Omega U^{\dagger}V \ .
\end{equation}
We can then compute, using the orthonormality of the spherical spinors \cite{SphericalSpinors} :
\begin{equation}
 \left(u_{L,p_L,j,\kappa,m_j}, u_{L,p'_L,j',\kappa',m_j'} \right)_t = \delta_{jj'} \delta_{\kappa \kappa'} \delta_{m_j m_j'} \int_0^{\infty} dR g^{\frac{3}{2}}(R) R^2 \left(\Phi_{L,p_L,j,\kappa} (R)\Phi_{L,p'_L,j',\kappa'} (R) +  \Psi_{L,p_L,j,\kappa} (R)\Psi_{L,p'_L,j',\kappa'} (R)\right) \ .
\end{equation}
We then impose the normalization
\begin{equation}\label{Eq.:Normalization}
 \int_0^{\infty} dR g^{\frac{3}{2}}(R) R^2 \left(\Phi_{L,p_L,j,\kappa} (R)\Phi_{L,p'_L,j',\kappa'} (R) +  \Psi_{L,p_L,j,\kappa} (R)\Psi_{L,p'_L,j',\kappa'} (R)\right) = \delta(p_L-p'_L) \ .
\end{equation}
The inner product of the antiparticle solutions is of course identical. The inner products among $p_L$ and $q_L$ solutions are required all to vanish, because they necessarily represent solutions with different energies. The analogous of Eq. \eqref{Eq.:Normalization} for $p_L \rightarrow q_L, p'_L \rightarrow q'_L$ is understood to hold as well.
The products between particle and antiparticle solutions all vanish due to the angular parts, except for
\begin{eqnarray}
 \nonumber \left(u_{L,p_L,j,\kappa,m_j}, v_{L,p'_L,j',\kappa',m_j'} \right)_t &=& ie^{i \left(E_L + E'_L \right)t} (-1)^{-m_j + \frac{3}{2}}\mathrm{sign}(\kappa')\delta_{jj'} \delta_{\kappa, - \kappa'} \delta_{m_j,- m_j'} \\
 &\times& \int_0^{\infty} dR g^{\frac{3}{2}}(R) R^2 \left(\Phi_{L,p_L,j,\kappa} (R)\Psi_{L,p'_L,j,-\kappa} (R) +  \Psi_{L,p_L,j,\kappa} (R)\Phi_{L,p'_L,j,-\kappa} (R)\right)
\end{eqnarray}
and the analogous $q_L,q'_L$ products where $E'_L = \sqrt{p^{'2}_L + M_L^2}$ ($E'_L= \sqrt{M_L^2 - q_L^{'2}}$ respectively). In order that $u,v$ constitute a complete orthornormal set of solutions, we must then require
\begin{equation}
 \int_0^{\infty} dR g^{\frac{3}{2}}(R) R^2 \left(\Phi_{L,p_L,j,\kappa} (R)\Psi_{L,p'_L,j,-\kappa} (R) +  \Psi_{L,p_L,j,\kappa} (R)\Phi_{L,p'_L,j,-\kappa} (R)\right) = 0
\end{equation}
for each $p_L, p'_L$, as well its $q_L$ counterpart. Note that this relation is enforced only for the inner product between modes of the same species (same $L$). The inner products between modes with different mass labels $L$ do instead lead to the Bogoliubov coefficients of mixing. We find explicitly
\begin{eqnarray}
 \nonumber \Xi_{p_1,j,\kappa,m_j; p'_2, j', \kappa', m_j'}(t) &=& \left(u_{1,p_1,j,\kappa,m_j}, u_{2,p'_2,j',\kappa',m_j'} \right)_t = e^{i \left(E_1 - E'_2\right)t}\delta_{jj'}\delta_{\kappa \kappa'} \delta_{m_j,m'_j} \delta (p_1 - p'_2) X_{p_1,j,\kappa}  \\
  \nonumber \Omega_{p_1,j,\kappa,m_j; p'_2, j', \kappa', m_j'}(t) &=& \left(u_{1,p_1,j,\kappa,m_j}, v_{2,p'_2,j',\kappa',m_j'} \right)_t = e^{i \left(E_1 + E'_2\right)t} \sign(\kappa) (-1)^{-m_j + \frac{1}{2}}\delta_{jj'}\delta_{\kappa,- \kappa'} \delta_{m_j,-m'_j} \delta (p_1 - p'_2) W_{p_1,j,\kappa} \\
\end{eqnarray}
where we have introduced the quantities
\begin{eqnarray}\label{Eq.:ReducedBogoliubov}
 \nonumber X_{p,j,\kappa} \delta (p-p') &=& \int_0^{\infty} dR g^{\frac{3}{2}}(R) R^2 \left[ \Phi_{1,p,j,\kappa} \Phi_{2,p',j,\kappa} +  \Psi_{1,p,j,\kappa} \Psi_{2,p',j,\kappa}   \right] \\
 W_{p,j,\kappa} \delta (p-p') &=& i\int_0^{\infty} dR g^{\frac{3}{2}}(R) R^2 \left[ \Phi_{1,p,j,\kappa} \Psi_{2,p',j,-\kappa} +  \Psi_{1,p,j,\kappa} \Phi_{2,p',j,-\kappa}   \right] \ .
\end{eqnarray}
Then, according to the general formulas \eqref{Eq.:FermionOperators}, the flavor operators are explicitly given by
\begin{eqnarray}\label{Eq.:SphericalFermionOperators}
 \nonumber a_{e,p,j,\kappa,m_j}(t) &=& \cos \Theta  \ a_{1,p,j,\kappa,m_j} + \sin \Theta \left[e^{i\left(E_1-E_2\right)t}X_{p,j,\kappa}a_{2,p,j,\kappa,m_j} + e^{i\left(E_1 + E_2\right)t}\mathrm{sign}(\kappa)(-1)^{-m_j + \frac{1}{2}}W_{p,j,\kappa}b^{\dagger}_{2,p,j,-\kappa,-m_j} \right]\\
 \nonumber a_{\mu,p,j,\kappa,m_j}(t) &=& \cos \Theta  \ a_{2,p,j,\kappa,m_j} - \sin \Theta \left[e^{i\left(E_2-E_1\right)t}X^*_{p,j,\kappa}a_{1,p,j,\kappa,m_j} + e^{i\left(E_1 + E_2\right)t}\mathrm{sign}(\kappa)(-1)^{-m_j + \frac{1}{2}}W_{p,j,-\kappa}b^{\dagger}_{1,p,j,-\kappa,-m_j} \right]\\
  \nonumber b^{\dagger}_{e,p,j,\kappa,m_j}(t) &=& \cos \Theta  \ b^{\dagger}_{1,p,j,\kappa,m_j} + \sin \Theta \left[e^{-i\left(E_1+E_2\right)t}\mathrm{sign}(\kappa)(-1)^{m_j + \frac{3}{2}}W^*_{p,j,\kappa}
  a_{2,p,j,-\kappa,-m_j} + e^{i\left(E_2 - E_1\right)t}X^*_{p,j,\kappa}b^{\dagger}_{2,p,j,\kappa,m_j} \right] \\
\nonumber b^{\dagger}_{\mu,p,j,\kappa,m_j}(t) &=& \cos \Theta  \ b^{\dagger}_{2,p,j,\kappa,m_j} - \sin \Theta \left[e^{-i\left(E_1+E_2\right)t}\mathrm{sign}(\kappa)(-1)^{m_j + \frac{3}{2}}W^*_{p,j,-\kappa}
  a_{1,p,j,-\kappa,-m_j} + e^{i\left(E_1 - E_2\right)t}X_{p,j,\kappa}b^{\dagger}_{1,p,j,\kappa,m_j} \right] \ . \\
\end{eqnarray}
Equations regarding the $q_L$ modes are obtained by the obvious replacements\footnote{It is understood that no mixing occurs between the $p_L$ and $q_L$ modes, which are mutually orthogonal even for different species $L,L'$.}.

\subsection{Components of the vacuum expectation value}

From the previous section, we realized that any spatial component of $\mathbb{T}_{\mu \nu}$ is zero. Let us now evaluate the remaining off-diagonal components $\mathbb{T}_{0i}$. It is sufficient to focus on just one of them, $\mathbb{T}_{0z}$, since by the underlying isotropy, the other components shall be identical. As convenient in this kind of explicit computations, we introduce an auxiliary tensor to keep track of the algebraic factors. We define for any two spinors $U,V$
\begin{equation}\label{Eq.:AuxiliaryTensor}
 B_{\mu \nu} (U,V) = \frac{i}{2} \left(\bar{U}\tilde{\gamma}_{(\mu} D_{\nu)}V - D_{(\mu} \bar{U} \tilde{\gamma}_{\nu)}V \right) \ .
\end{equation}
By inserting the field expansions \eqref{Eq.:FreeFieldExpansion} with respect to the modes of Eqs. \eqref{Eq.:SolutionAnsatz} and \eqref{Eq.:AntiparticleSolution} into Eq. \eqref{Eq.:EnergyMomentumTensor},  we get
\begin{eqnarray}
 \nonumber :T_{\mu \rho}: &=& \sum_{L=1,2}\sum_{j,j',\kappa,\kappa',m_j,m_j'}\int_0^{\infty} dp \int_0^{\infty}dp' \Bigg \lbrace B_{\mu \rho} \left( u_{L,p,j,\kappa,m_j},u_{L,p',j',\kappa',m_j'} \right)a^{\dagger}_{L,p,j,\kappa,m_j}a_{L,p',j',\kappa',m_j'} \\
 \nonumber &+& B_{\mu \rho} \left( u_{L,p,j,\kappa,m_j},v_{L,p',j',\kappa',m_j'} \right)a^{\dagger}_{L,p,j,\kappa,m_j}b^{\dagger}_{L,p',j',\kappa',m_j'} + B_{\mu \rho} \left( v_{L,p,j,\kappa,m_j},u_{L,p',j',\kappa',m_j'} \right)b_{L,p,j,\kappa,m_j}a_{L,p',j',\kappa',m_j'} \\
 &-& B_{\mu \rho} \left( v_{L,p,j,\kappa,m_j},v_{L,p',j',\kappa',m_j'} \right)b^{\dagger}_{L,p',j',\kappa',m_j'}b_{L,p,j,\kappa,m_j}  + (p \rightarrow q)\ .
\end{eqnarray}
The last term represents the contribution of  $q$ ($E_L \leq M_L$) modes, that is formally identical, except possibly for the range of pseudo-momenta $q$, which is limited respectively by $M_1$ and $M_2$, for the $L=1$ and $L=2$ contributions.
The expectation values on the flavor vacuum can be computed by the usual trick:
\begin{eqnarray}
 \nonumber  \bra{0_F(t)} a^{\dagger}_{L,p,j,\kappa,m_j}a_{L,p',j',\kappa',m_j'} \ket{0_F(t)} &=& \bra{0}\mathcal{J}_{\Theta} (t) a^{\dagger}_{L,p,j,\kappa,m_j}a_{L,p',j',\kappa',m_j'} \mathcal{J}^{-1}_{\Theta}(t) \ket{0}  \\
\nonumber &=&  \bra{0}\mathcal{J}_{\Theta} (t) a^{\dagger}_{L,p,j,\kappa,m_j}\mathcal{J}^{-1}_{\Theta}\mathcal{J}_{\Theta} (t)a_{L,p',j',\kappa',m_j'} \mathcal{J}^{-1}_{\Theta}(t) \ket{0}  \\
\nonumber &=& \bra{0}\mathcal{J}^{-1}_{-\Theta} (t) a^{\dagger}_{L,p,j,\kappa,m_j}\mathcal{J}_{-\Theta}\mathcal{J}^{-1}_{-\Theta} (t)a_{L,p',j',\kappa',m_j'} \mathcal{J}_{-\Theta}(t) \ket{0} \\
&=& \bra{0}  a^{\dagger}_{L,p,j,\kappa,m_j}(-\Theta) a_{L,p',j',\kappa',m_j'} (-\Theta) \ket{0} \ ,
\end{eqnarray}
where $a_{L,p,j,\kappa,m_j}(-\Theta)$ are the transformed operators of Eq. \eqref{Eq.:SphericalFermionOperators} for $\Theta \rightarrow - \Theta$ and it is understood that $1 \rightarrow e, 2 \rightarrow \mu$. The results are
\begin{eqnarray}
 \nonumber \bra{0_F(t)} a^{\dagger}_{L,p,j,\kappa,m_j}a_{L,p',j',\kappa',m_j'} \ket{0_F(t)} &=& \delta(p-p') \delta_{jj'}\delta_{\kappa \kappa'} \delta_{m_j m_j'} \sin^2 \Theta |W_{p,j,k}|^2 = \bra{0_F(t)} b^{\dagger}_{L,p',j',\kappa',m_j'}b_{L,p,j,\kappa,m_j} \ket{0_F(t)} \\
 \nonumber \bra{0_F(t)} a^{\dagger}_{1,p,j,\kappa,m_j}b_{1,p',j',\kappa',m_j'} \ket{0_F(t)} &=& \delta (p-p') \delta_{jj'}\delta_{\kappa,-\kappa'}\delta_{m_j, -m_j'} \sin^2 \Theta e^{-2iE_1t} \mathrm{sign}(\kappa) (-1)^{-m_j + \frac{1}{2}}W^*_{p,j,\kappa}X^*_{p,j,-\kappa} \\
\nonumber \bra{0_F(t)} a^{\dagger}_{2,p,j,\kappa,m_j}b_{2,p',j',\kappa',m_j'} \ket{0_F(t)} &=& \delta (p-p') \delta_{jj'}\delta_{\kappa,-\kappa'}\delta_{m_j, -m_j'} \sin^2 \Theta e^{-2iE_2t} \mathrm{sign}(\kappa) (-1)^{-m_j + \frac{1}{2}}W^*_{p,j,-\kappa}X_{p,j,-\kappa} \ . \\
\end{eqnarray}
The remaining expectation values can be obtained by complex conjugation and relabeling. The VEV of the energy-momentum tensor, on the flavor vacuum, is consequently
\begin{eqnarray}\label{Eq.:ExplicitVEV}
 \nonumber \mathbb{T}_{\mu \rho} &=& \sin^2 \Theta \sum_{j,\kappa,m_j}\int_0^{\infty}dp \Bigg \lbrace  |W_{p,j,\kappa}|^2 \sum_{L=1,2}\left[B_{\mu \rho} \left( u_{L,p,j,\kappa,m_j},u_{L,p,j,\kappa,m_j}\right) - B_{\mu \rho} \left( v_{L,p,j,\kappa,m_j},v_{L,p,j,\kappa,m_j} \right) \right] \\
\nonumber &+& \left(B_{\mu \rho} \left( u_{1,p,j,\kappa,m_j},v_{1,p,j,-\kappa,-m_j}\right)e^{-2iE_1t}\mathrm{\sign}(\kappa)(-1)^{-m_j + \frac{1}{2}}W^*_{p,j,\kappa}X^*_{p,j,-\kappa} + \mathrm{c.c.} \right) \\
&+& \left(B_{\mu \rho} \left( u_{2,p,j,\kappa,m_j},v_{2,p,j,-\kappa,-m_j}\right)e^{-2iE_2t}\mathrm{\sign}(\kappa)(-1)^{-m_j + \frac{1}{2}}W^*_{p,j,-\kappa}X_{p,j,-\kappa} + \mathrm{c.c.} \right) \bigg \rbrace  + (p \rightarrow q) \ .
\end{eqnarray}
As pertaining $\mathbb{T}_{0z}$, we only need a reduced part of $B_{0z}(U,V)$, namely
\begin{equation}
 B_{0z}(U,V) \equiv \frac{i}{2} \left( \bar{U} \tilde{\gamma}_z D_0 V - D_0 \bar{U} \tilde{\gamma}_z V\right)
\end{equation}
because the other terms vanish in the expectation value, by the same reasoning leading to Eq. \eqref{Eq.:NoSpatialComponents}. Let us first evaluate $B_{0 z} \left( u_{L,p,j,\kappa,m_j},u_{L,p,j,\kappa,m_j}\right)$. The covariant time derivative yields
\begin{equation}\label{Right}
 D_0 u_{L,p,j,\kappa,m_j} = e^{-iE_Lt}\begin{pmatrix}
                                       \left(-iE_L \Phi_{L,p,j,\kappa} - i \frac{f'(R)}{4\sqrt{f(R)g(R)}}\Psi_{L,p,j,\kappa} \right) \chi_{\kappa, m_j} \\
                                       \left(E_L \Psi_{L,p,j,\kappa} -  \frac{f'(R)}{4\sqrt{f(R)g(R)}}\Phi_{L,p,j,\kappa} \right) \chi_{-\kappa, m_j}
                                      \end{pmatrix} \ ,
\end{equation}
so that it does not change the angular structure of $u_{L,p,j,\kappa,m_j}$, only its radial part is modified. The same holds for $D_0 v_{L,p,j,\kappa,m_j}$ and for the adjoints. On the other hand, it is
\begin{equation}\label{Left}
\bar{u}_{L,p,j,\kappa,m_j} \tilde{\gamma}_z = - \sqrt{g(R)} e^{iE_Lt} \begin{pmatrix}
          - i \Psi_{L,p,j,\kappa}\left(G_{-\kappa,m_j}\chi^{\dagger}_{-\kappa,m_j} + H_{-\kappa,m_j}\chi^{\dagger}_{\kappa-1,m_j} \right)   \\
          \Phi_{L,p,j,\kappa}\left(G_{\kappa,m_j}\chi^{\dagger}_{\kappa,m_j} + H_{\kappa,m_j}\chi^{\dagger}_{-\kappa-1,m_j} \right) \ ,
                                                                      \end{pmatrix}^{T}
\end{equation}
where $T$ stands for matrix transpositon and we have used the properties of Eq. \eqref{Eq.:SphericalSpinorProperties}. The action of $\tilde{\gamma}_z$ changes the angular structure, in particular modifying the $\kappa$ index of the spinors. Due to the orthogonality of the spherical spinors for different $\kappa$, the contraction between Eqs. \eqref{Left} and \eqref{Right} vanishes. Since a similar conclusion is valid also for $D_0 \bar{u}_L \tilde{\gamma}_z u_L$, we conclude that $B_{0 z} \left( u_{L,p,j,\kappa,m_j},u_{L,p,j,\kappa,m_j}\right) = 0$.

Notice now that the antiparticle solutions with opposite angular quantum numbers, namely $v_{L,p,j,-\kappa,-m_j}$
have the same angular structure as $u_{L,p,j,\kappa,m_j}$, as evident from Eq. \eqref{Eq.:AntiparticleSolution}. Therefore any combination of $B_{0z} (u,v), B_{0z}(v,v)$ appearing in $\mathbb{T}_{0z}$ is zero, due to the mismatch in the angular parts. It  leads to the result
\begin{equation}
 \mathbb{T}_{0z} = 0 \ .
\end{equation}
By isotropy, the only non vanishing component of the VEV is $\mathbb{T}_{00}$. Following the same steps outlined above, the energy density is found to be
\begin{eqnarray}\label{Eq.:ExplicitEnergyDensity}
 \nonumber \mathbb{T}_{00} &=&  4 \sin^2 \Theta \sqrt{f(R)} \sum_{L=1,2} \sum_{j,\kappa,m_j}\int_0^{\infty} dp \ E_L |W_{p,j,\kappa}|^2 \left(|\Phi_{L,p,j,\kappa}|^2 + |\Psi_{L,p,j,\kappa}|^2 \right)  \\
 &+& 4 \sin^2 \Theta \sqrt{f(R)} \sum_{L=1,2} \sum_{j,\kappa,m_j}\int_0^{M_L} dq \ E_L |W_{q,j,\kappa}|^2 \left(|\Phi_{L,q,j,\kappa}|^2 + |\Psi_{L,q,j,\kappa}|^2 \right)\ .
\end{eqnarray}
We recall that for the $q$ contribution\footnote{If the spectrum of ``bound'' states $E_L \leq M_L$ is not continuous, the integral in the last term is to be replaced by a summation.}, we have $E_L = \sqrt{M_L^2 - q^2}$. Note that $\mathbb{T}_{00} > 0$, it depends only on $R$ and it is formally infinite, requiring regularization. Due to its proportionality to $\sin^2 \Theta$, this energy density is clearly zero in absence of mixing.

\subsection{The Minkowskian limit and weak field approximation}

The solution to the Dirac equation, or equivalently to the system of equations \eqref{Eq.:RadialSystem} for the radial functions, represents a remarkable challenge for generic metric functions $f(R), g(R)$. We now turn to the flat spacetime limit, both for a check of the general formalism and to gain valuable insight on the possible behavior of $\mathbb{T}_{00}$.

\vspace{0.3cm}
\emph{- Minkowskian Limit} - In flat space the metric functions become constants $f(R),g(R) \rightarrow 1$, and the system of equations for the radial functions is extremely simplified:
\begin{eqnarray}\label{Eq.:FlatRadialSystem}
 \nonumber \left(\partial_R + \frac{1-\kappa}{R} \right)\Psi^0_{L,p,j,\kappa}(R) &=& \left(M_L  - E_L \right) \Phi^0_{L,p,j,\kappa}(R) \\
 \left(\partial_R + \frac{1+\kappa}{R} \right)\Phi^0_{L,p,j,\kappa}(R) &=& \left(M_L  + E_L \right) \Psi^0_{L,p,j,\kappa}(R)  \ .
\end{eqnarray}
Here we have added the apex $0$ to signal these are the flat space radial functions. The two equations can be combined to yield a second order differential equation for $\Psi$
\begin{equation}
\frac{d^2 \Psi^0_{L,p,j,\kappa}}{dR^2} + \frac{2}{R}\frac{d \Psi^0_{L,p,j,\kappa}}{dR} - \frac{\kappa(\kappa-1)}{R^2}\Psi^0_{L,p,j,\kappa} + p^2 \Psi^0_{L,p,j,\kappa} = 0
\end{equation}
where as usual $p^2 = E_L^2-M_L^2$. By the substitution $z=pR$, we recognize the spherical Bessel equation
\begin{equation}
 z^2 \frac{d^2 \Psi^0_{L,p,j,\kappa}}{dz^2} + 2z\frac{d \Psi^0_{L,p,j,\kappa}}{dz} + \left[z^2 - \kappa(\kappa-1) \right]\Psi^0_{L,p,j,\kappa} = 0
\end{equation}
for which the spherical bessel functions $j_{\kappa-1}(pR)$ and $y_{\kappa-1}(pR)$ are independent solutions \cite{Abramowitz}. For $\kappa \geq 1$, the functions $j_{\kappa -1}(pR)$ are regular as $R \rightarrow 0$, while  $y_{\kappa -1}(pR) = (-1)^{\kappa}j_{-\kappa}(pR)$ are singular and must therefore be discarded.

The situation is reversed for $\kappa  < 1$. We then have
\begin{equation}
  \Psi^0_{L,p,j,\kappa}(R)=\begin{cases}
    N_{L,p,j,\kappa} \ j_{\kappa -1 }(pR) & \text{if $\kappa \geq 1$}.\\
    N'_{L,p,j,\kappa} \ y_{\kappa -1 }(pR) & \text{otherwise},
  \end{cases}
\end{equation}
where $N_{L,p,j,\kappa}$ and $N'_{L,p,j,\kappa}$ are (possibly distinct) normalization constants to determine. By inverting the first of Eq. \eqref{Eq.:FlatRadialSystem} we find for $\Phi$
\begin{equation}
  \Phi^0_{L,p,j,\kappa}(R)= \frac{p}{M_L-E_L}\begin{cases}
    N_{L,p,j,\kappa} \ j_{\kappa }(pR) & \text{if $\kappa \geq 1$}.\\
    N'_{L,p,j,\kappa} \ y_{\kappa}(pR) & \text{otherwise}.
  \end{cases}
\end{equation}
Next we determine the normalization constants by imposing Eq. \eqref{Eq.:Normalization}, with the aid of the orthonormality relations among the spherical Bessel functions \cite{Abramowitz}. The final result is
\begin{eqnarray}
\nonumber  \Psi^0_{L,p,j,\kappa}(R)&=&p (M_L - E_L) \frac{1}{\sqrt{\pi E_L(E_L - M_L)}}\begin{cases}
     \ j_{\kappa -1 }(pR) & \text{if $\kappa \geq 1$}.\\
    \ y_{\kappa -1 }(pR) & \text{otherwise}.
    \end{cases} \\
    \Phi^0_{L,p,j,\kappa}(R)&=&p^2 \frac{1}{\sqrt{\pi E_L(E_L - M_L)}}\begin{cases}
     \ j_{\kappa}(pR) & \text{if $\kappa \geq 1$}.\\
    \ y_{\kappa  }(pR) & \text{otherwise}.
    \end{cases} \ .
\end{eqnarray}
The corresponding energy density
\begin{equation}\label{Eq.:FlatDensity}
 \mathbb{T}_{00}^{(0)} = 4 \sin^2 \Theta \sum_{L=1,2} \sum_{j,\kappa,m_j}\int_0^{\infty} dp \ E_L |W^0_{p,j,\kappa}|^2 \left(|\Phi^0_{L,p,j,\kappa}|^2 + |\Psi^0_{L,p,j,\kappa}|^2 \right) =: 4 \sin^2 \Theta  \ \mathcal{K}
\end{equation}
has been evaluated in \cite{Capolupo2016} and it turns out to be a spacetime constant\footnote{Note that in the flat spacetime limit the isotropic coordinates here employed reduce to the standard cartesian coordinates.}, which we here denote by $4 \sin^2 \Theta  \ \mathcal{K} $.
The constant $\mathcal{K}$ reads explicitly \cite{Capolupo2016}:
\begin{equation}
 \frac{\Delta M}{8\pi^2} \sum_{L=1,2} (-1)^L \left[\Lambda M_L \omega_L (\Lambda) - M_L^3 \ln \left(\frac{\Lambda + \omega_L(\Lambda)}{M_L} \right) \right]
\end{equation}
where $\Delta M = M_2 - M_1, \omega_L (p) = \sqrt{p^2 + M_L^2}$ and $\Lambda$ is an ultraviolet momentum cutoff.

\vspace{0.3cm}
\emph{ - Weak field approximation -}
Let us now consider the weak field form of the metric, with $f(R) = 1 + 2 V(R)$ and $g(R) = 1 - 2V(R)$. The potential $V(R)$ is regarded as a small correction, and only terms at most linear in $V(R)$ are kept in every physical quantity. If the flavor vacuum is the only energy-momentum source, the potential satisfies the Poisson equation
\begin{equation}\label{Eq.:PoissonEquation}
 \nabla^2 V = 4 \pi G \mathbb{T}_{00} \ .
\end{equation}
Determining the exact solutions to the Dirac equations for this metric, and therefore $\mathbb{T}_{00}$, even specifying $V(R)$, proves to be an exceptionally difficult task. Rather than solving the Dirac equation directly, we proceed as follows. Let us start with the flat space solutions $\Phi^0, \Psi^0$ and ask what multiplicative factor $h(R)$ can modify them in order to keep the normalization \eqref{Eq.:Normalization}. To be precise, we assume
\begin{equation}
 \Phi_{L,p,j,\kappa} (R) = h(R)  \Phi^0_{L,p,j,\kappa}(R) \ ; \ \ \ \ \ \Psi_{L,p,j,\kappa} (R) = h(R)  \Psi^0_{L,p,j,\kappa}(R)
\end{equation}
and compute
\begin{eqnarray}
\nonumber && \int_0^{\infty} dR g^{\frac{3}{2}}(R) R^2 \left(\Phi_{L,p_L,j,\kappa} (R)\Phi_{L,p'_L,j',\kappa'} (R) +  \Psi_{L,p_L,j,\kappa} (R)\Psi_{L,p'_L,j',\kappa'} (R)\right) \\
&& \simeq \int_0^{\infty} dR (1-3V(R)) R^2 h^2(R) \left(\Phi^0_{L,p_L,j,\kappa} (R)\Phi^0_{L,p'_L,j',\kappa'} (R) +  \Psi^0_{L,p_L,j,\kappa} (R)\Psi^0_{L,p'_L,j',\kappa'} (R)\right)
\end{eqnarray}
where $g^{\frac{3}{2}}(R)$ has been expanded to first order in $V(R)$. If $h(R)$ is chosen so to compensate for the extra potential term $(1-3V(R))h^2(R) \stackrel{!}{=} 1 + O(V^2)$, namely $h(R) = \left(1 + \frac{3}{2}V(R)\right)$, it follows that
\begin{eqnarray}
\nonumber && \int_0^{\infty} dR g^{\frac{3}{2}}(R) R^2 \left(\Phi_{L,p_L,j,\kappa} (R)\Phi_{L,p'_L,j',\kappa'} (R) +  \Psi_{L,p_L,j,\kappa} (R)\Psi_{L,p'_L,j',\kappa'} (R)\right) \\
&& \simeq \int_0^{\infty} dR  R^2  \left(\Phi^0_{L,p_L,j,\kappa} (R)\Phi^0_{L,p'_L,j',\kappa'} (R) +  \Psi^0_{L,p_L,j,\kappa} (R)\Psi^0_{L,p'_L,j',\kappa'} (R)\right)
\end{eqnarray}
and normalization \eqref{Eq.:Normalization} is automatically preserved by these approximate solutions up to linear order in $V(R)$. We then set
\begin{equation}\label{Eq.:ApproximateSolutions}
 \Phi_{L,p,j,\kappa} (R) = \left(1 + \frac{3}{2} V(R) \right)  \Phi^0_{L,p,j,\kappa}(R) \ ; \ \ \ \ \ \Psi_{L,p,j,\kappa} (R) = \left(1 + \frac{3}{2} V(R) \right)  \Psi^0_{L,p,j,\kappa}(R) \ .
\end{equation}
We further assume that the potential is sufficiently weak that no $q$ states, i.e. with $E_L \leq M_L$ are formed, so that the second contribution in Eq. \eqref{Eq.:ExplicitEnergyDensity} vanishes just as in flat space. In this circumstance the modes of Eq. \eqref{Eq.:ApproximateSolutions}, satisfying Eq. \eqref{Eq.:Normalization} constitute an orthonormal and complete set for the field expansion. It can be checked immediately from Eq. \eqref{Eq.:ReducedBogoliubov} that the Bogoliubov coefficients equal the flat space ones up to linear order in $V(R)$, $W_{p,j,\kappa} = W^0_{p,j,k} + O(V^2)$. As a consequence the energy density reads
\begin{eqnarray}\label{Eq.:ApproximateDensity}
 \nonumber && \mathbb{T}_{00} = 4 \sin^2 \Theta \sqrt{f(R)} \sum_{L=1,2} \sum_{j,\kappa,m_j}\int_0^{\infty} dp \ E_L |W_{p,j,\kappa}|^2  h^2(R)\left(|\Phi^0_{L,p,j,\kappa}|^2 + |\Psi^0_{L,p,j,\kappa}|^2 \right) \\
 \nonumber &\simeq& 4 \sin^2 \Theta \left(1+ V(R) \right)\left(1+3V(R) \right) \sum_{L=1,2} \sum_{j,\kappa,m_j}\int_0^{\infty} dp \ E_L |W^0_{p,j,\kappa}|^2  \left(|\Phi^0_{L,p,j,\kappa}|^2 + |\Psi^0_{L,p,j,\kappa}|^2 \right) \\
 &\simeq& 4 \sin^2 \Theta \ \mathcal{K} \left(1 + 4 V(R) \right)
\end{eqnarray}
where the first equality follows from expanding up to linear order in $V(R)$ and the second stems from Eq. \eqref{Eq.:FlatDensity}. Remarkably, the result of Eq. \eqref{Eq.:ApproximateDensity} allows us to solve for $V(R)$ in the Poisson equation. Inserting Eq. \eqref{Eq.:ApproximateDensity} in Eq. \eqref{Eq.:PoissonEquation}, we obtain, with $\alpha= 16\pi G \mathcal{K} \sin^2 \Theta (> 0)$,
\begin{equation}
 V'' + \frac{2}{R}V' - \alpha (1 + 4V) = 0 \ .
\end{equation}
This can be reduced to a simple form by the substitution $Z = R(1+4V)$, which gives
\begin{equation}
 Z'' - 4\alpha Z = 0.
\end{equation}
The positive root must be discarded for a sensible flat space limit for $R\rightarrow \infty$. The potential is then, up to an irrelevant additive  constant \footnote{One simply shifts the zero of the potential.},
\begin{equation}\label{Eq.:Yukawa}
 V(R) = C \frac{e^{-2\sqrt{\alpha}R}}{R} \ .
\end{equation}
Here $C$ is an integration constant with dimensions of length. The gravitational potential due to the flavor vacuum has the form of a Yukawa potential.

\begin{table}[h]
\caption{Data for spiral galaxies, extrapolated from \cite{DataSpirals}. The third column is the total baryonic mass (stars plus gas) in $10^{10}$ solar masses $M_{\odot}$. }
\begin{center}\label{SpiralData}
\begin{tabular}{c c c}
 \hline \hline
 Galaxy  & $v_{FLAT}$ & $\mathcal{M}$  \\ [0.5ex]
  & ($\mathrm{km/s}$) & ($10^{10} M_{\odot}$) \\
 \hline
 UGC 2885 & 300 & 35.8  \\
 \
 NGC 2841 & 287  & 34  \\
 NGC 5533 & 250 & 22 \\
 NGC 6674 & 242 & 21.9 \\
 NGC 3992 & 242 & 16.22 \\
 NGC 7331 & 232 & 14.4 \\
 NGC 3953 & 223 & 8.17 \\
 NGC 5907 & 214 & 10.8 \\
 NGC 2998 & 213 & 11.3 \\
 NGC 801 & 208 & 12.9 \\
 NGC 5371 & 208 & 12.5 \\
 NGC 5033 & 195 & 9.73 \\
 NGC 3893 & 188 & 4.76 \\
 NGC 4157 & 185 & 5.62 \\
 NGC 2903 & 185 & 5.81 \\
 NGC 4217 & 178 & 4.5 \\
 NGC 4013 & 177 & 4.84 \\
 NGC 3521 & 175 & 7.13 \\
 NGC 4088 & 173 & 4.09 \\
 NGC 3877 & 167 & 3.49 \\
 NGC 4100 & 164 & 4.62 \\
 NGC 3949 & 164 & 1.72 \\
 NGC 3726 & 162 & 3.24 \\
 NGC 6946 & 160 & 5.4 \\
 NGC 4051 & 159 & 3.29 \\
 NGC 3198 & 156 & 2.93 \\
 NGC 2683 & 155 & 3.55 \\
 NGC 3917 & 135 & 1.58 \\
 NGC 4085 & 134 & 1.13 \\
 NGC 2403 & 134 & 1.57 \\
 [1ex]
 \hline
 \hline
\end{tabular}
\quad
\begin{tabular}{c c c}
 \hline \hline
 Galaxy  & $v_{FLAT}$ & $\mathcal{M}$  \\ [0.5ex]
  & ($\mathrm{km/s}$) & ($10^{10} M_{\odot}$) \\
 \hline
 NGC 3972 & 134 & 1.12 \\
 UGC 128 & 131 & 1.48 \\
 NGC 4010 & 128 & 1.13 \\
 F568-VI & 124 & 1 \\
 NGC 3769 & 122 & 1.33 \\
 NGC 6503 & 121 & 1.07 \\
 F568-3 & 120 & 0.83 \\
 NGC 4183 & 112 & 0.93 \\
 F563-V2 & 111 & 0.87 \\
 F563-1 & 111 & 0.79 \\
 NGC 1003 & 110 & 1.12 \\
 UGC 6917 & 110 & 0.74 \\
 UGC 6930 & 110 & 0.73 \\
 M33 & 107 & 0.61 \\
 UGC 6983 & 107 & 0.86 \\
 NGC 247 & 107 & 0.53 \\
 NGC 7793 & 100 & 0.51 \\
 NGC 300 & 90 & 0.35 \\
  NGC 5585 & 90 & 0.37 \\
 NGC 55 & 86 & 0.23 \\
 UGC 6667 & 86 & 0.33 \\
 UGC 2259 & 86 & 0.27 \\
 UGC 6446 & 82 & 0.42 \\
 UGC 6818 & 73 & 0.14 \\
 NGC 1560 & 72 & 0.132 \\
 IC 2574 & 66 & 0.077 \\
 DDO 170 & 64 & 0.085 \\
 NGC 3109 & 62 & 0.073 \\
 DDO 154 & 56 & 0.049 \\
 DDO 168 & 54 & 0.037 \\
 [1ex]
 \hline
 \hline
 \end{tabular}
\end{center}
\end{table}

\begin{table}[h]
\caption{Data for gas rich galaxies, extrapolated from \cite{DataGas}. The third column is the total baryonic mass (stars plus gas) in $10^{10}$ solar masses $M_{\odot}$. }
\begin{center}\label{GasData}
\begin{tabular}{c c c}
 \hline \hline
 Galaxy  & $v_{FLAT}$ & $\mathcal{M}$  \\ [0.5ex]
  & ($\mathrm{km/s}$) & ($10^{10} M_{\odot}$) \\
 \hline
 NGC 3198 & 149 & 3.26  \\
 NGC 2403 & 134  & 0.996  \\
 F568-V1 & 124 & 0.686 \\
 F568-3 & 120 & 0.929 \\
 F568-1 & 118 & 1.057 \\
 F563-V2 & 111 & 0.684 \\
 UGC 6983 & 108 & 0.888 \\
 UGC 3711 & 95 & 0.185 \\
 NGC 5585 & 90 & 0.282 \\
 NGC 2915 & 84 & 0.07 \\
 IC 2233 & 84 & 0.3 \\
 F571-V1 & 83 & 0.314 \\
 F565-V2 & 83 & 0.129 \\
 UGC 5721 & 79 & 0.127 \\
 UGC 8490 & 78 & 0.114 \\
 NGC 1560 & 77 & 0.22 \\
 UGC 4499 & 74 & 0.265 \\
 UGC 6818 & 72 & 0.356 \\
 IC 2574 & 68 & 0.246 \\
 D 500-2 & 68 & 0.117 \\
 UGC 8055 & 66 & 0.117 \\
 NGC 3109 & 66 & 0.064 \\
 UGC 9211 & 64 & 0.175 \\
 UGC 3851 & 60 & 0.151 \\
 [1ex]
 \hline
 \hline
\end{tabular}
\quad
\begin{tabular}{c c c}
 \hline \hline
 Galaxy  & $v_{FLAT}$ & $\mathcal{M}$  \\ [0.5ex]
  & ($\mathrm{km/s}$) & ($10^{10} M_{\odot}$) \\
 \hline
 UGC 4115 & 59 & 0.044 \\
 D 575-2 & 59 & 0.046 \\
 UGC 8550 & 58 & 0.047 \\
 KKH 11 & 56 & 0.009 \\
 DDO 168 & 54 & 0.067 \\
 D 631-7 & 53 & 0.02 \\
 D 500-3 & 45 & 0.0037 \\
 NGC 3741 & 44 & 0.03 \\
 UGC 6145 & 41 & 0.0052 \\
 UGC 7242 & 40 & 0.0097 \\
 KK 98251 & 38 & 0.013 \\
 UGCA 444 & 38 & 0.0078 \\
 D 512-2 & 37 & 0.013 \\
 UGCA 92 & 37 & 0.027 \\
 KK 9824 & 35 & 0.014 \\
 P 51659 & 31 & 0.0075 \\
 DDO 181 & 30 & 0.0054 \\
 D 564-8 & 29 & 0.0027 \\
 UGC 8833 & 27 & 0.0029 \\
 DDO 183 & 25 & 0.0052 \\
 UGC 8215 & 20 & 0.0035 \\
 CamB & 20 & 0.0031 \\
 DDO210 & 17 & 0.0005 \\
 [1ex]
 \hline
 \hline
 \end{tabular}
\end{center}
\end{table}

\begin{table}
\caption{Best fit values for $\beta$ and $\nu$ for the two galaxy datasets of tables \ref{SpiralData} and \ref{GasData}.}
\begin{center}\label{Table1}
\begin{tabular}{c  c  c  c}
 \hline
 Fit parameter & \ \ Spiral galaxies (\cite{DataSpirals} and Table \ref{SpiralData}) \ \ & \ \ Gas dominated galaxies (\cite{DataGas} and Table \ref{GasData}) & Combined data \ \ \\ [0.5ex]
 \hline
 $\beta$ & 0.433117 & 0.417015 & 0.426762 \\
 \hline
 $\nu$ & -0.520693  & -0.517175 & -0.515629 \\ [1ex]
 \hline
\end{tabular}
\end{center}
\end{table}

\begin{table}
\caption{Best fit values for $\beta_0$ for two different cutoffs, $ \Lambda_0 \simeq 1.5425 \ \mathrm{keV}$ and $\Lambda_{EW} = 246 \mathrm{GeV}$. The same data as in Table \ref{Table1} is used as input.}
\begin{center}\label{Table2}
\begin{tabular}{c c c c}
 \hline
 Fit parameter & \ \ Spiral galaxies (\cite{DataSpirals} and Table \ref{SpiralData}) \ \ & \ \ Gas dominated galaxies (\cite{DataGas} and Table \ref{GasData}) \ \ & Combined data \\ [0.5ex]
 \hline
 $\beta_0$ at $\Lambda = \Lambda_0$ & 0.413572 & 0.422315 & 0.414944 \\
 \hline
 $\beta_0$ at $\Lambda = \Lambda_{EW}$ & $2.59436 \times 10^{-8}$  & $2.64805 \times 10^{-8}$ & $2.60183 \times 10^{-8}$ \\ [1ex]
 \hline
\end{tabular}
\end{center}
\end{table}

\section{GALACTIC DYNAMICS FROM NEUTRINO MIXING}

The above considerations, in particular the final result,  can be applied to the dynamics of spiral galaxies. In particular, we will show that both flat rotation curves and the baryonic Tully-Fisher relation can be achieved in this framework.

\subsection{Flat rotation curves of galaxies}

As discussed for instance in \cite{Jusufi, Cardone, Stabile}, a Yukawa form of the gravitational potential might explain the flat rotation curves of galaxies. For a given galaxy with total baryonic mass $\mathcal{M}$, the gravitational potential is
\begin{equation}
 V(R) = V_{B} (R) + V_{F} (R)
\end{equation}
where $V_B(R) = - \frac{G\mathcal{M}}{R}$ is the standard Newtonian potential due to the baryonic matter and $V_F(R)$ is the potential due to the flavor vacuum of Eq. \eqref{Eq.:Yukawa}. By setting the integration constant in Eq. \eqref{Eq.:Yukawa} equal to $C = -G\mathcal{M} \beta$, with $\beta$ a freely specifiable dimensionless parameter, and further introducing $d = \frac{1}{2 \sqrt{\alpha}}$, we have
\begin{equation}\label{Eq.:TotalPotential}
 V(R) = -\frac{G\mathcal{M}}{R}\left( 1 + \beta e^{-\frac{R}{d}}\right) \ .
\end{equation}
A test mass $m$ is subject to the force $-m \nabla V(R)$, so that, for circular orbits, the velocity $v(R)$ has to satisfy
\begin{equation}
 \frac{v^2(R)}{R} =  \frac{dV}{dR} \ .
\end{equation}
It follows, from Eq. \eqref{Eq.:TotalPotential},  that
\begin{equation}\label{Eq.:CircularVelocity}
 v^2 (R) = \frac{G\mathcal{M}}{R}  \left[1 + \beta e^{-\frac{R}{d}}\left(1+ \frac{R}{d} \right) \right] \ .
\end{equation}
Expanding the exponential at first order,  one has approximately
\begin{equation}
 v^2(R) \simeq \frac{G\mathcal{M}}{R} + \frac{G\mathcal{M}\beta (R + d)}{Rd} \ .
\end{equation}
Towards the outer region of the galaxy, the first term becomes negligible and the second approaches the constant value
\begin{equation}
 v^2_{\mathrm{FLAT}}(R)\simeq \frac{G\mathcal{M}\beta}{d}
\end{equation}
so that, in principle, the profile of Eq. \eqref{Eq.:CircularVelocity} can explain the flatness of the rotation curves. It has indeed been proven that the Yukawa potential can satisfactorily reproduce the rotation curves of the Milky Way and M31 galaxies \cite{CapozzielloJusufi} once the bulge and disk structures are appropriately taken into account. According to the best fit values presented in \cite{CapozzielloJusufi} both $\beta$ and $d$ depend on the galaxy under consideration, in particular $\beta_{MW} = 0.4, d_{MW}= 0.74 \ \mathrm{kpc}$ and $\beta_{M31} = 0.37, d_{M31} = 0.52 \ \mathrm{kpc}$. For a discussion about the variability of $d$ on cosmological scales, see \cite{CapozzielloJusufi}. For later usage, we take the total baryonic mass of the Milky to be \cite{MilkyWayMass,CapozzielloJusufi} $\mathcal{M}_{MW} \simeq 1.2 \times 10^{40} \ \mathrm{kg}$.

\begin{figure}[h]
\begin{centering}
\includegraphics[scale=0.43]{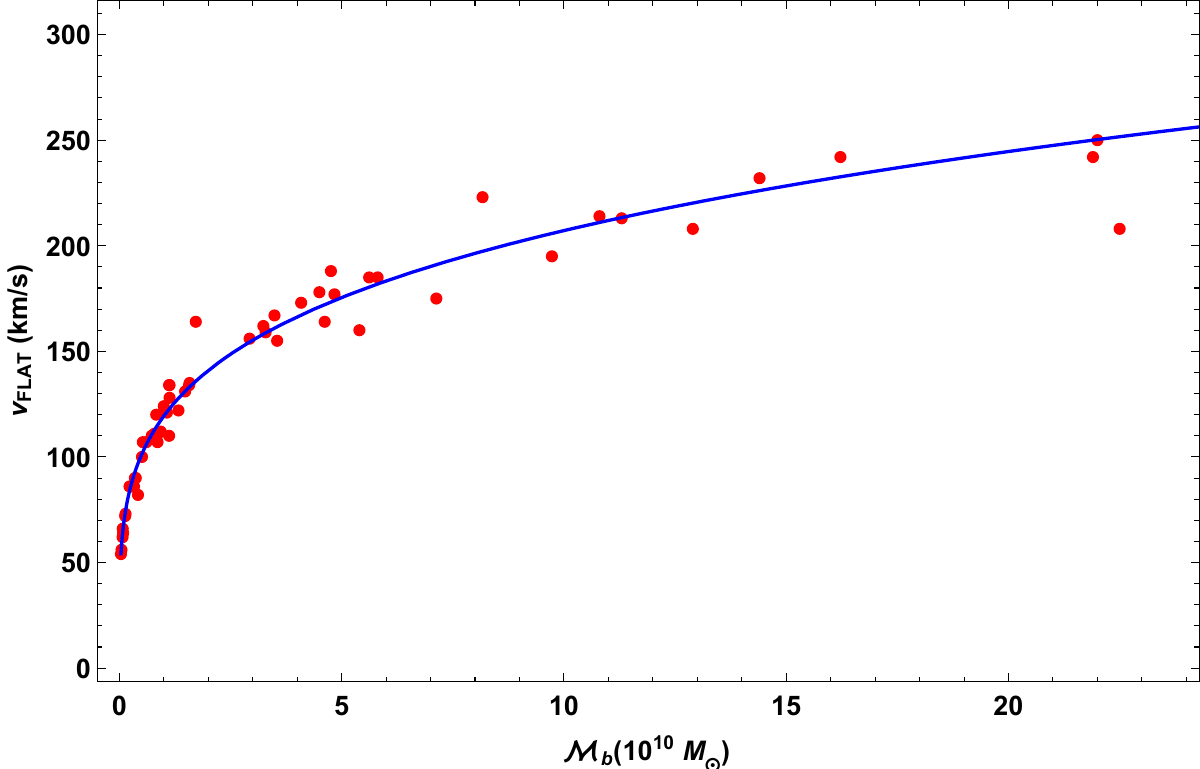} \hfill \includegraphics[scale=0.43]{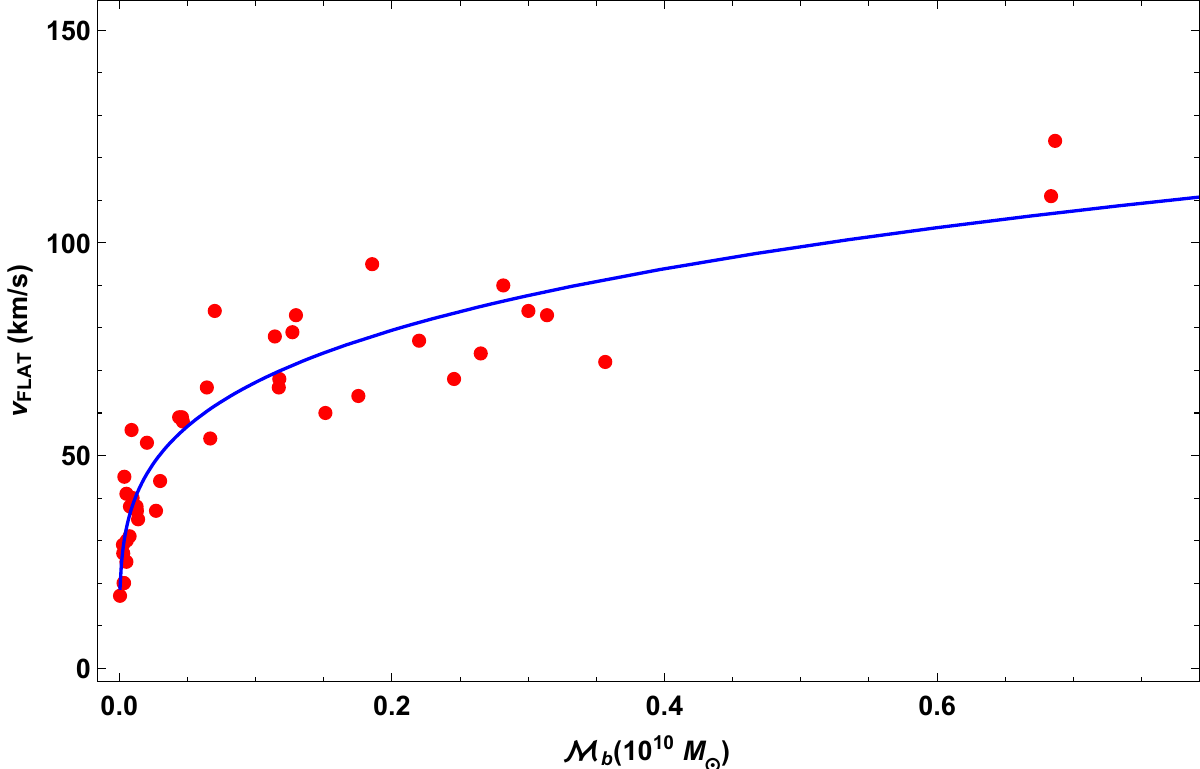}
\includegraphics[scale=0.43]{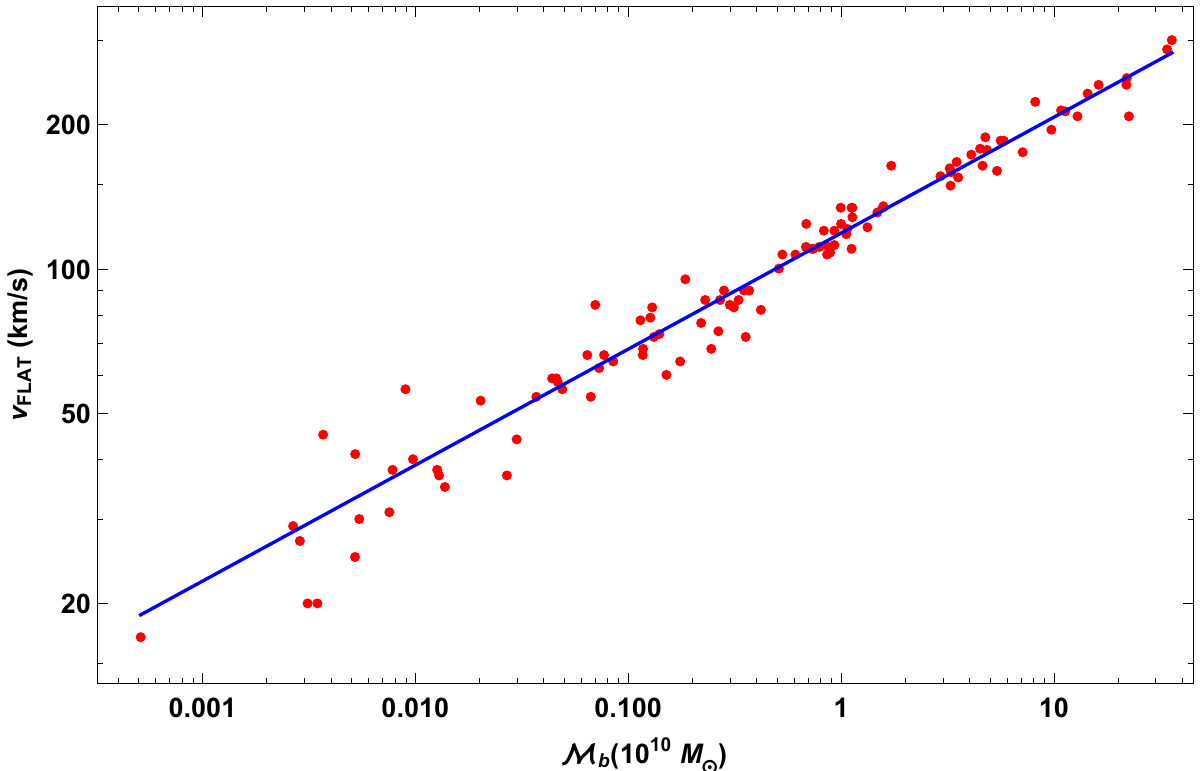}
\par\end{centering}
\caption{\label{Figura1} (color online) Plots of the best fit (solid blue line) for Eq. \eqref{BTF} for constant $\beta$ and cutoff scaling $\Lambda = \Lambda_0 \left(\frac{\mathcal{M}}{\mathcal{M}_{MW}} \right)^{\nu}$ versus the experimental points (red): (top left) spiral galaxies dataset (Table \ref{SpiralData}), (top right) gas rich galaxies dataset (Table \ref{GasData}), (bottom) logarithmic scale plot for the combined dataset. For the best fit parameters see Table \ref{Table1}.}
\end{figure}

\begin{figure}[h]
\begin{centering}
\includegraphics[scale=0.43]{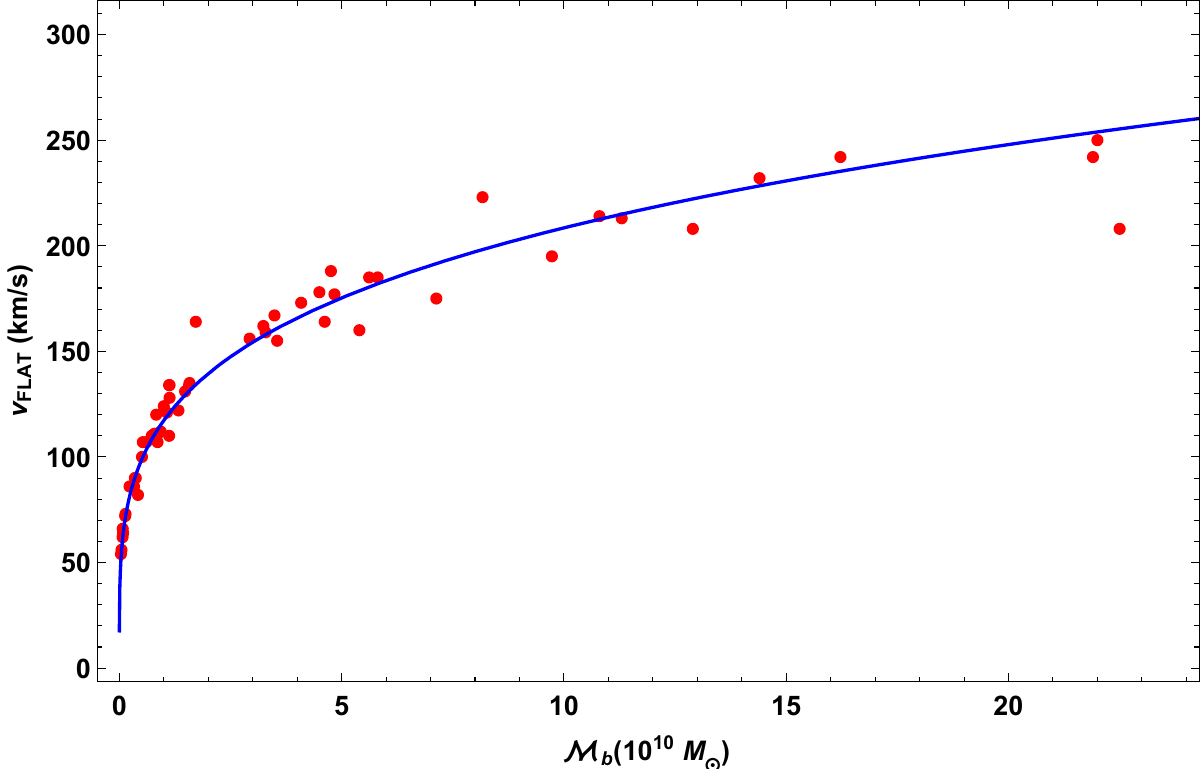} \hfill \includegraphics[scale=0.43]{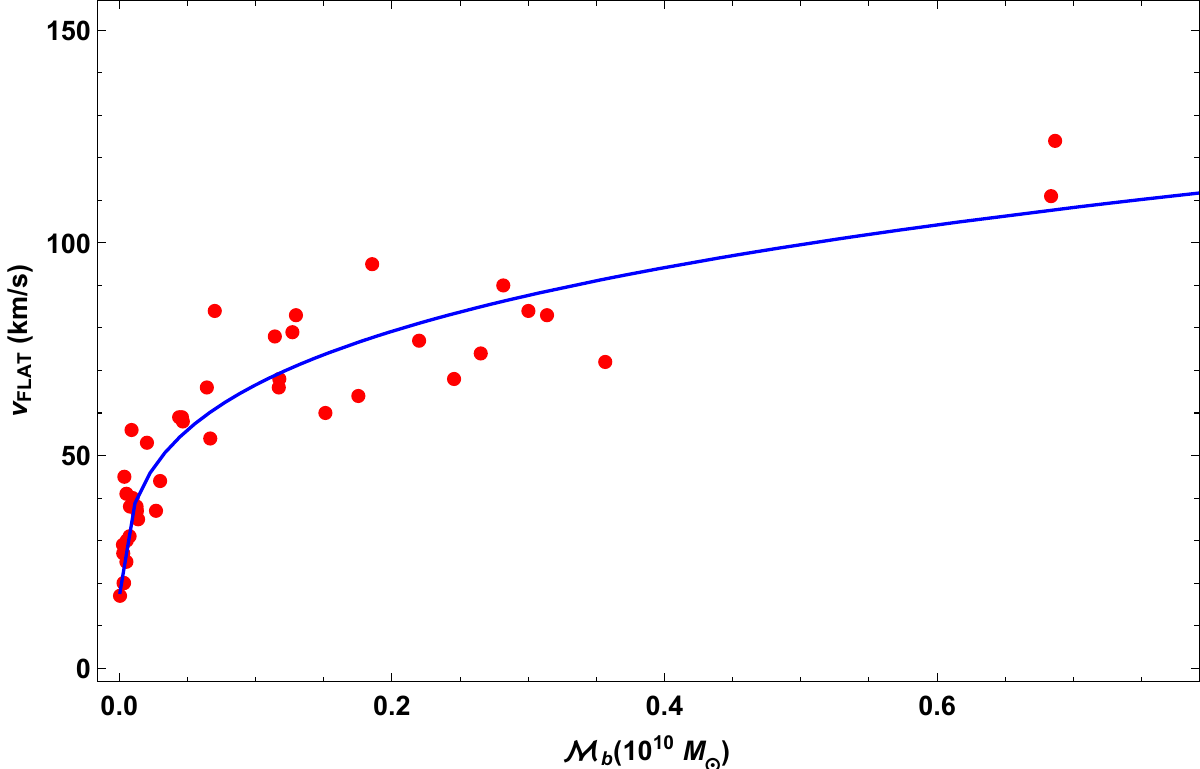}
\includegraphics[scale=0.43]{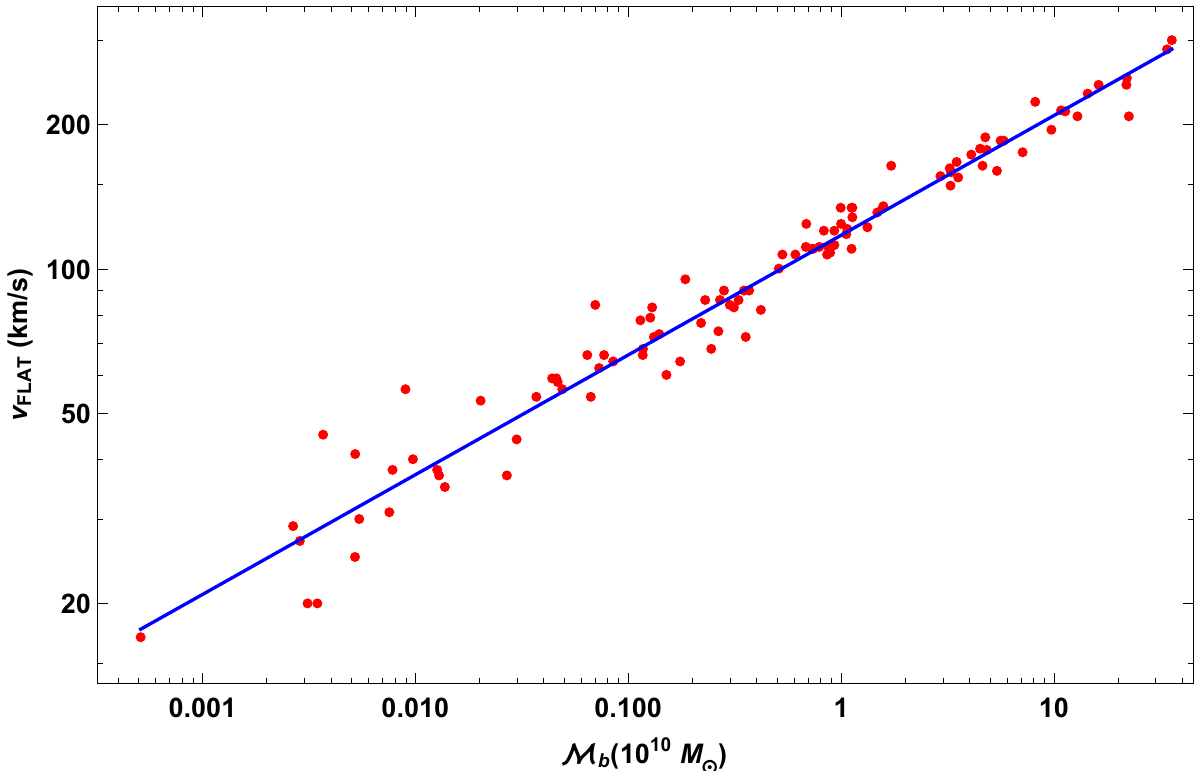}
\par\end{centering}
\caption{\label{Figura2} (color online) Plots of the best fit (solid blue line) for Eq. \eqref{BTF} for constant cutoff  $\Lambda = \Lambda_0 \simeq \ 1.5425 \ \mathrm{keV}$ and $\beta$ scaling as $\beta = \beta_0 \left(\frac{\mathcal{M}}{\mathcal{M}_{MW}} \right)^{-\frac{1}{2}}$ versus the experimental points (red): (top left) spiral galaxies dataset (Table \ref{SpiralData}), (top right) gas rich galaxies dataset (Table \ref{GasData}), (bottom) logarithmic scale plot for the combined dataset. For the best fit parameters see Table \ref{Table2}.}
\end{figure}

\begin{figure}[h]
\begin{centering}
\includegraphics[scale=0.7]{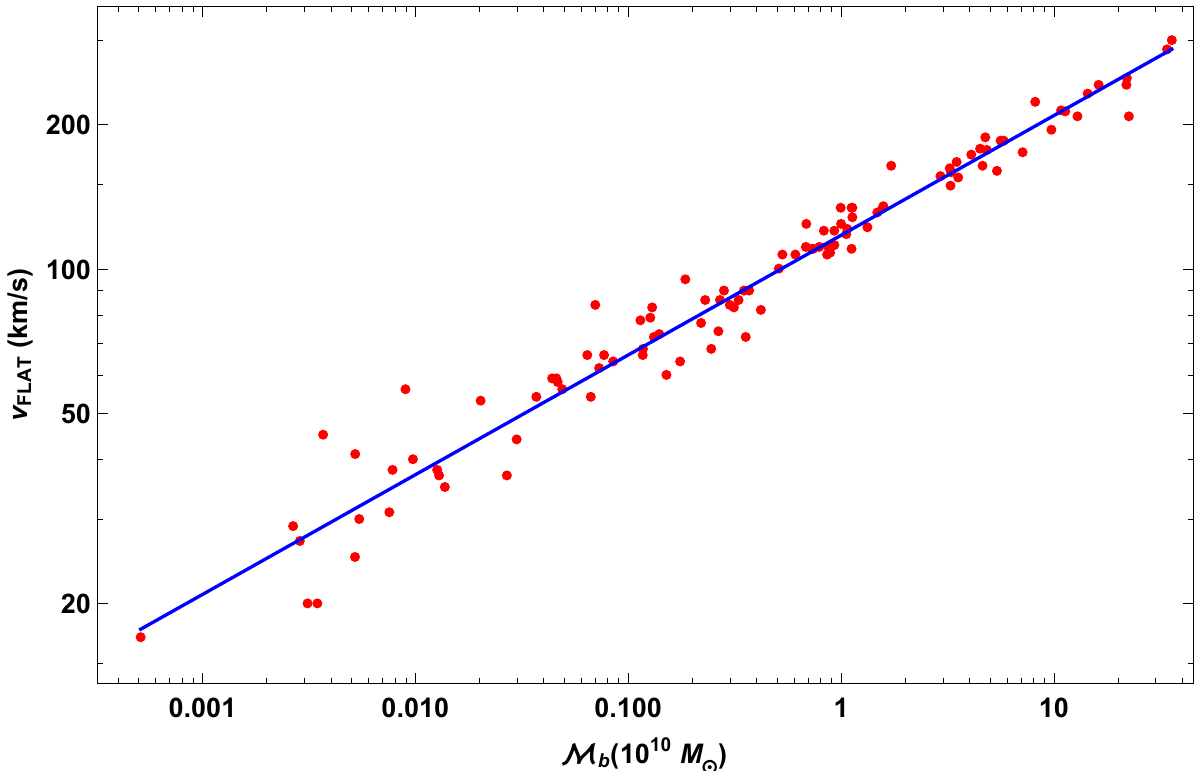}
\par\end{centering}
\caption{\label{Figura3} (color online) Logarithmic scale plot of the best fit (solid blue line) for Eq. \eqref{BTF} for constant cutoff  $\Lambda = \Lambda_{EW} \simeq \ 246 \ \mathrm{GeV}$ and $\beta$ scaling as $\beta = \beta_0 \left(\frac{\mathcal{M}}{\mathcal{M}_{MW}} \right)^{-\frac{1}{2}}$ versus the experimental points (red). The experimental points are from the combined dataset and the best fit parameter is shown in Table \ref{Table2}.}
\end{figure}

\subsection{Baryonic Tully-Fisher relation}

Another desirable feature of the Yukawa potential is its capability to reproduce the Milgrom result $v_{FLAT}^4 = GM a_0$ \cite{Milgrom, Milgrom2}, where $a_0$ is a constant acceleration and $v_{FLAT}$ denotes the circular velocity in the flat area of the rotation curve \cite{CapozzielloJusufi,Jusufi}. In the outer region of the galaxy we indeed have
\begin{equation}
 v^4_{FLAT} (R) \simeq G\mathcal{M}\left(\frac{G\mathcal{M}(R+d)^2 \beta^2}{R^2 d^2} \right)e^{-\frac{2R}{d}}
\end{equation}
so that setting
\begin{equation}\label{YukawaAcceleration}
 a_{0} = \lim_{R\rightarrow d} G\mathcal{M}\left(\frac{G\mathcal{M}(R+d)^2 \beta^2}{R^2 d^2} \right)e^{-\frac{R}{d}} = \frac{4G\mathcal{M} \beta^2 e^{-2}}{d^2}
\end{equation}
we have
\begin{equation}
  v^4_{FLAT} \simeq G\mathcal{M} a_0 \ .
\end{equation}
Since the Yukawa acceleration of Eq. \eqref{YukawaAcceleration} has to be a universal constant $a_0 \simeq 1.2 \times 10^{-10} \mathrm{m s^{-2}}$, it is clear that either $\beta$ or $d$, or even both, have to vary with the galaxy size (i.e., at this level, they need to depend on the total baryonic mass $\mathcal{M}$). The analysis in \cite{CapozzielloJusufi} seems to favor a nearly universal value for $\beta$ and a scale dependent $d$. Recalling that $d \propto \alpha^{-\frac{1}{2}} \sim \Lambda^{-1}$, where $\Lambda$ is the ultraviolet cutoff of Eq. \eqref{Eq.:FlatDensity}, the ultraviolet cutoff has to scale with the total baryonic mass. It is immediate to see, from Eq. \eqref{YukawaAcceleration},  that one should indeed have $\Lambda \sim \mathcal{M}^{-\frac{1}{2}}$.
There is however a second possibility for the baryonic Tully-Fisher relation to be reproduced with a constant ultraviolet cutoff and an appropriate scaling of $\beta \sim \mathcal{M}^{-\frac{1}{2}}$. Exquisitely from the point of view of QFT this latter possibility is more appealing.

We now consider two sets of galaxy data, one for spirals \cite{DataSpirals} and another one for gas dominated galaxies \cite{DataGas}, fitting the baryonic Tully Fisher relation
\begin{equation}\label{BTF}
 v^4_{FLAT} = \xi \mathcal{M}
\end{equation}
with $\xi = G a_0$ as provided by the Yukawa potential of Eq. \eqref{Eq.:TotalPotential} according to Eq. \eqref{YukawaAcceleration}.
We take into account the two aforementioned scenarios:
\begin{enumerate}
\item Universal $\beta$ and varying cutoff, with $\Lambda = \Lambda_0 \left(\frac{\mathcal{M}}{\mathcal{M}_{MW}} \right)^{\nu}$. Here $\Lambda_0$ is picked to reproduce the value $d_{MW} = 0.74 \ \mathrm{kpc}$ found in \cite{CapozzielloJusufi} and equals $\Lambda_0 \simeq 1.5425 \ \mathrm{keV}$, while $\nu$ is an exponent to be determined. The results are presented in Fig. \ref{Figura1} with the best fit values summarized in Table \ref{Table1}. The data agrees with $\beta \simeq 0.42$ and $\nu \simeq -0.52$, in accordance with the expected approximate $\mathcal{M}^{-\frac{1}{2}}$ scaling of the cutoff.

\item Universal $d$ and universal cutoff, varying $\beta$, with $\beta = \beta_0 \left(\frac{\mathcal{M}}{\mathcal{M}_{MW}} \right)^{-\frac{1}{2}}$ and $\beta_0$ a parameter to be determined. The results are presented in Figs. \ref{Figura2} and \ref{Figura3} with the best fit values summarized in table \ref{Table2}. For $\Lambda = \Lambda_0$, $\beta_0 \simeq \  0.42$, while it is much smaller for larger cutoff, $\beta_0 \simeq 2.6 \times 10^{-8}$ for $\Lambda = \Lambda_{EW} = 246 \ \mathrm{GeV}$.

\end{enumerate}

It should be noted that Eq. \eqref{YukawaAcceleration} is insensitive to the sign of $\beta$, so that also $-\beta$ of Table \ref{Table1} and $-\beta_0$ of Table \ref{Table2} are equally valid best fit values.

\section{Discussion and Conclusions}

Using QFT in curved space we have derived general results pertaining the flavor vacuum and its associated energy-momentum tensor. We have shown that in most cases of interest the latter attains a perfect fluid form, characterized by the dust and cold dark matter equation of state. We have then delved into the important case of static spherically symmetric spacetimes, deriving an explicit expression for the energy density due to the flavor vacuum. In the weak field approximation, we have solved the Poisson equation with the above mentioned source, and found that it is solved by a Yukawa form of the potential. Borrowing from results in the recent literature pertaining Yukawa cosmology, we have argued that such potential can explain the flatness of the rotation curves of spiral galaxies. In this way the flavor vacuum plays the role of (part of?) the dark matter component. By leveraging on the theoretical Tully-Fisher relation that emerges from the Yukawa potential, akin to the MOND result $v^4_{FLAT} = G\mathcal{M} a_0$, we have fitted the experimental baryonic Tully-Fisher relation for two important datasets, including gas rich galaxies. We have discussed two possible scenarios, one for which the ultraviolet cutoff for the energy density is held constant and one for which the cutoff scales with the galaxy size. Both scenarios are able to reproduce the experimental baryonic Tully-Fisher relation and then a possible MOND behavior for galactic dynamics as for extended theories of gravity \cite{DataGas, Tula}. In future works the analysis may be refined beyond the approximations used, aiming at exact solutions to the Dirac equations and the full non-linear Poisson equation.

\section*{Acknowledgements}
Authors  thank R. Serao for his useful comments and supportive attitude.
They  acknowledge partial financial support from MUR and  Istituto Nazionale di Fisica Nucleare (INFN), Sezione di Napoli and Gruppo Collegato di Salerno,  Iniziative Specifiche QGSKY and MoonLight-2.
 A.C. also acknowledges the COST Action CA1511 Cosmology and Astrophysics Network for
Theoretical Advances and Training Actions (CANTATA);
S.C. also acknowledges the COST Action CA21136 (CosmoVerse).

\end{document}